\documentstyle{arPrep}

\newcommand\lesssim{\mbox{$^{<}\hspace{-0.24cm}_{\sim}$}}
\newcommand\gtrsim{\mbox{$^{>}\hspace{-0.24cm}_{\sim}$}}
\def\msun{$M_{\odot}$}

\def\ergsec{\hbox{erg s$^{-1}$ }}

\def\RS{$R_{\rm S}$}

\def\fcol{$f_{\rm col}$}
\def\xte{{RXTE\/} }
 
\def \araa{{\it Annu. Rev. Astron. Astrophys.\/} }
\def \apj{{\it Ap.~J.\/} }
\def \apjs{{\it Ap.~J. Supp.\/} }

\def \aanda{{\it Astron. Astrophys.\/} }

\def \mnras{{\it MNRAS\/} }
\def \pasj{{\it PASJ\/} }

\begin{document}

\input epsf.tex
\input epsf.def
\input psfig.sty

\jname{Annual Reviews of Astronomy \& Astrophysics}
\jyear{2005}
\jvol{44}
\ARinfo{}

\title{X-ray Properties of Black-Hole Binaries}

\markboth{Remillard \& McClintock}{Black-Hole Binaries}

\author{Ronald A. Remillard 
\affiliation{Kavli Institute for Astrophysics and Space Research, MIT,
Cambridge, Massachusetts 02139 email: rr@space.mit.edu}
Jeffrey E. McClintock
\affiliation{Harvard-Smithsonian Center for Astrophysics, Cambridge, Massachusetts 02138
email:jem@cfa.harvard.edu }}


\begin{abstract}
We review the properties and behavior of 20 X-ray binaries that
contain a dynamically-confirmed black hole, 17 of which are transient
systems.  During the past decade, many of these transient sources were
observed daily throughout the course of their typically year-long
outburst cycles using the large-area timing detector aboard the {\it
Rossi X-ray Timing Explorer}.  The evolution of these transient
sources is complex.  Nevertheless, there are behavior patterns common
to all of them as we show in a comprehensive comparison of six
selected systems.  Central to this comparison are three X-ray states
of accretion, which are reviewed and defined quantitatively.  We
discuss phenomena that arise in strong gravitational fields, including
relativistically-broadened Fe lines, high-frequency quasi-periodic
oscillations (100-450 Hz), and relativistic radio and X-ray jets.
Such phenomena show us how a black hole interacts with its
environment, thereby complementing the picture of black holes that
gravitational wave detectors will provide.  We sketch a scenario for
the potential impact of timing/spectral studies of accreting black
holes on physics and discuss a current frontier topic, namely, the
measurement of black hole spin.
\end{abstract}

\maketitle


\section{Introduction}

Oppenheimer \& Snyder (1939) made the first rigorous calculation
describing the formation of a black hole (BH).  The first strong
evidence for such an object came from X-ray and optical observations
of the X-ray binary Cygnus X--1 (Bolton 1972; Webster \& Murdin 1972).
Today, a total of 20 similar X-ray binary systems are known that
contain a compact object believed to be too massive to be a neutron
star or a degenerate star of any kind (i.e., $M > 3$ \msun).  These
systems, which we refer to as black-hole binaries (BHBs), are the
focus of this review.

These 20 dynamical BHs are the most visible representatives of an
estimated $10^8$--$10^9$ stellar-mass BHs that are believed to exist in
the Galaxy (e.g., Brown \& Bethe 1994; Timmes, Woosley, \& Weaver
1996).  Stellar-mass BHs are important to astronomy in numerous ways.
For example, they are one endpoint of stellar evolution for massive
stars, and the collapse of their progenitor stars enriches the
universe with heavy elements (Woosley et al.\ 2002).  Also, the
measured mass distribution for even the small sample featured here is
used to constrain models of BH formation and binary evolution (e.g.,
Fryer \& Kalogera 2001; Podsiadlowski et al.\ 2003). Lastly, some BHBs
appear to be linked to the hypernovae believed to power gamma-ray
bursts (Israelian et al.\ 1999; Brown et al.\ 2000; Orosz et al.\
2001).

The BHBs featured here are mass-exchange binaries that contain an
accreting BH primary and a nondegenerate secondary star. For
background on X-ray binaries, see Psaltis (2006).  For comprehensive
reviews on BHBs, see McClintock \& Remillard (2006),
Tanaka and Shibazaki (1996) and Tanaka \& Lewin (1995).  X-ray
observations of BHBs allow us to gain a better understanding of BH
properties and accretion physics.  In this review, we emphasize those
results that challenge us to apply the predictions of general
relativity (GR) in strong gravity (\S8). Throughout, we make extensive
use of the extraordinary data base amassed since January 1996 by
NASA's {\it Rossi X-ray Timing Explorer} (RXTE; Swank 1998).

In an astrophysical environment, a BH is completely specified in GR by
two numbers, its mass $M$ and its specific angular momentum or spin $a
= J / c M$, where $J$ is the BH angular momentum and $c$ is the speed
of light. The spin value is conveniently expressed in terms of a
dimensionless spin parameter, $a_{*} = a / R_{\rm g}$, where the
gravitational radius is $R_{\rm g} \equiv GM/c^{2}$. The mass simply
supplies a scale, whereas the spin changes the geometry.  The value of
$a_{*}$ lies between 0 for a Schwarzschild hole and 1 for a
maximally-rotating Kerr hole.  A defining property of a BH is its
event horizon, the immaterial surface that bounds the interior region
of space-time that cannot communicate with the external universe.  The
event horizon, the existence of an innermost stable circular orbit
(ISCO), and other properties of BHs are discussed in many texts (e.g.,
Shapiro \& Teukolsky 1983; Kato et al.\ 1998).  The radius of the
event horizon of a Schwarzschild BH ($a_{*} = 0$) is \RS = 2$R_{\rm
g}$ = 30~km($M/$10\msun), the ISCO lies at $R_{\rm ISCO}$~=~6$R_{\rm
g}$, and the corresponding maximum orbital frequency is $\nu_{\rm
ISCO}=220$~Hz($M/$10\msun)$^{-1}$.  For an extreme Kerr BH ($a_{*} =
1$), the radii of both the event horizon and the ISCO (prograde
orbits) are identical, $R_{\rm K}~=~R_{\rm ISCO}$~=~$R_{\rm g}$, and
the maximum orbital frequency is $\nu_{\rm ISCO}~=~
1615$~Hz($M/$10\msun)$^{-1}$.

\section{A Census of Black-Hole Binaries and Black-Hole Candidates}

Following the discovery of Cygnus X--1, the second BHB to be
identified was LMC X--3 (Cowley et al.\ 1983).  Both sources are
persistently bright in X-rays, and their secondaries are massive
O/B-type stars (White et al.\ 1995).  The third identified BHB,
A~0620--00, is markedly different (McClintock \& Remillard 1986).
A~0620--00 was discovered as an X-ray nova in 1975 when it suddenly
brightened to an intensity of 50 Crab\footnote{1~Crab~=~
$2.43\times10^{-9}$~erg~cm$^{-2}$~s$^{-1}$~keV$^{-1}$ =~1.00~mJy~
(averaged over 2--11 keV) for a Crab-like spectrum with photon index
$\Gamma=2.08$; Koyama et al.\ (1984).} to become the brightest
nonsolar X-ray source ever observed (Elvis et al.\ 1975).  Then, over
the course of a year, the X-ray nova decayed back into quiescence to
become a feeble (1~{\it $\mu$}Crab) source (McClintock et al.\ 1995).
Similarly, the optical counterpart faded from outburst maximum by
$\Delta V\approx7.1$ mags to $V\approx18.3$ in quiescence, thereby
revealing the optical spectrum of a K-dwarf secondary.

As of this writing, there are a total of 20 confirmed BHBs and,
remarkably, 17 of them are X-ray novae like A~0620--00.  They are
ordered in the top half of Table~1 by right ascension (column~1).
Column~2 gives the common name of the source (e.g., LMC X--3) {\it or}
the prefix to the coordinate name that identifies the discovery
mission (e.g., XTE J, where a ``J'' indicates that the coordinate
epoch is J2000).  For X-ray novae, the third column gives the year of
discovery and the number of outbursts that have been observed.  The
spectral type of the secondary star is given in column~4. Extensive
optical observations of this star yield the key dynamical data
summarized respectively in the last three columns: the orbital period,
the mass function, and the BH mass.  Additional data on BHBs are given
in Tables 4.1 \& 4.2 of McClintock \& Remillard (2006).

An  observational quantity  of special  interest is  the  mass function,
$f(M)\equiv  P_{\rm  orb}K_{2}^{3}/2\pi G=M_{1}$sin$^3i/(1+q)^{2}$  (see
Table~1, column  6).  The observables on the left  side of this equation
are  the orbital  period $P_{\rm  orb}$  and the  half-amplitude of  the
velocity curve of the secondary  $K_{2}$.  On the right, the quantity of
greatest interest is $M_{1}$, the mass of the BH primary (given  in column~7);  the other
parameters are the  orbital inclination angle $i$ and  the mass ratio $q
\equiv M_{2}/M_{1}$,  where $M_{2}$ is  the mass of the  secondary.  The
value  of  $f(M)$ can  be  determined  by  simply measuring  the  radial
velocity curve of the secondary star, and it corresponds to the absolute
minimum allowable mass of the compact object.

\begin{table}%
\small
\caption{Twenty confirmed  black holes and twenty black hole candidates$^a$}
\begin{tabular}{llllccc}
\toprule
Coordinate       &Common$^b$  &Year$^c$  &Spec.    &P$_{\rm orb}$ &f(M)          &M$_{1}$ \\
Name             &Name/Prefix &         &         &(hr)          &(M$_{\odot})$ &(M$_{\odot})$ \\
\colrule
0422+32          &(GRO~J)     &1992/1   &M2V      &5.1   &1.19$\pm$0.02   &3.7--5.0   \\
0538--641        &LMC~X--3    & --      &B3V      &40.9  &2.3$\pm$0.3     &5.9--9.2    \\
0540--697        &LMC~X--1    & --      &O7III    &93.8$^d$  &0.13$\pm0.05^d$   &4.0--10.0:$^e$ \\
0620--003        &(A)         &1975/1$^f$   &K4V      &7.8   &2.72$\pm$0.06   &8.7--12.9   \\
1009--45         &(GRS)       &1993/1   &K7/M0V   &6.8   &3.17$\pm$0.12   &3.6--4.7:$^e$    \\
1118+480         &(XTE J)     &2000/2   &K5/M0V   &4.1   &6.1$\pm$0.3     &6.5--7.2    \\
1124--684        &Nova Mus 91 &1991/1   &K3/K5V   &10.4  &3.01$\pm$0.15   &6.5--8.2    \\
1354--64$^g$     &(GS)        &1987/2   &GIV      &61.1$^g$  &5.75$\pm$0.30   & --     \\
1543--475        &(4U)        &1971/4   &A2V      &26.8  &0.25$\pm$0.01   &8.4--10.4   \\
1550--564        &(XTE~J)     &1998/5   &G8/K8IV  &37.0  &6.86$\pm$0.71   &8.4--10.8   \\
1650--500$^h$    &(XTE~J)     &2001/1   &K4V      &7.7   &2.73$\pm$0.56   & --         \\
1655--40         &(GRO~J)     &1994/3   &F3/F5IV  &62.9  &2.73$\pm$0.09   &6.0--6.6    \\
1659--487        &GX~339--4  &1972/10$^i$ &  --   &42.1$^{j,k}$ &5.8$\pm$0.5    & --   \\
1705--250        &Nova Oph 77 &1977/1   &K3/7V    &12.5  &4.86$\pm$0.13   &5.6--8.3    \\
1819.3--2525     &V4641 Sgr   &1999/4   &B9III    &67.6  &3.13$\pm$0.13   &6.8--7.4    \\
1859+226         &(XTE~J)     &1999/1   &  --     &9.2:$^e$ &7.4$\pm$1.1:$^e$  &7.6--12.0:$^e$ \\
1915+105         &(GRS)     &1992/Q$^l$  &K/MIII  &804.0 &9.5$\pm$3.0     &10.0--18.0  \\
1956+350         &Cyg~X--1    & --      &O9.7Iab  &134.4 &0.244$\pm$0.005 &6.8--13.3   \\
2000+251         &(GS)        &1988/1   &K3/K7V   &8.3   &5.01$\pm$0.12   &7.1--7.8    \\
2023+338         &V404 Cyg  &1989/1$^f$ &K0III    &155.3 &6.08$\pm$0.06   &10.1--13.4  \\
\hline
1524--617        &(A)         & 1974/2  & --      & --   & --             & --         \\
1630--472        &(4U)        & 1971/15 & --      & --   & --             & --         \\
1711.6--3808     &(SAX~J)     & 2001/1  & --      & --   & --             & --         \\
1716--249        &(GRS)       & 1993/1  & --      &14.9  & --             & --         \\
1720--318        &(XTE~J)     & 2002/1  & --      & --   & --             & --         \\
1730--312        &(KS)        & 1994/1  & --      & --   & --             & --         \\
1737--31         &(GRS)       & 1997/1  & --      & --   & --             & --         \\
1739--278        &(GRS)       & 1996/1  & --      & --   & --             & --         \\
1740.7--2942     &(1E)        & --      & --      & --   & --             & --         \\
1743--322        &(H)         & 1977/4  & --      & --   & --             & --         \\
1742--289        &(A)         & 1975/1  & --      & --   & --             & --         \\
1746--331        &(SLX)       & 1990/2  & --      & --   & --             & --         \\
1748--288        &(XTE~J)     & 1998/1  & --      & --   & --             & --         \\
1755--324        &(XTE~J)     & 1997/1  & --      & --   & --             & --         \\
1755--338        &(4U)     & 1971/Q$^l$  & --     &4.5   & --             & --         \\
1758--258        &(GRS)    & 1990/Q$^l$  & --     & --   & --             & --         \\
1846--031        &(EXO)       & 1985/1  & --      & --   & --             & --         \\
1908+094         &(XTE~J)     & 2002/1  & --      & --   & --             & --         \\
1957+115         &(4U)        & --      & --      &9.3   & --             & --         \\
2012+381         &(XTE~J)     & 1998/1  & --      & --   & --             & --         \\
\botrule
\multicolumn{7}{l}{$^a$See McClintock \& Remillard (2006; and references therein) for columns 3--5, Orosz (2003) for columns 6--7,} \\
\multicolumn{7}{l}{plus additional references given below.} \\
\multicolumn{7}{l}{$^b$A prefix to a coordinate name is enclosed in parentheses.  The presence/absence of a} \\
\multicolumn{7}{l}{``J''indicates that the epoch of the coordinates is J2000/B1950.} \\
\multicolumn{7}{l}{$^c$Year of initial X-ray outburst/total number of X-ray outbursts.} \\
\multicolumn{7}{l}{$^d$Period and f(M) corrections by AM Levine and D Lin, private communication.} \\
\multicolumn{7}{l}{$^e$Colon denotes uncertain value or range.} \\
\multicolumn{7}{l}{$^f$Additional outbursts in optical archives: A~0620 (1917) and V404 Cyg (1938, 1956).} \\
\multicolumn{7}{l}{$^g$Casares et al.\ 2004; possible alias period of 61.5 hr.} \\
\multicolumn{7}{l}{$^h$Orosz et al.\ 2004.} \\
\multicolumn{7}{l}{$^i$Estimated by Kong et al.\ 2002.} \\
\multicolumn{7}{l}{$^j$Hynes et al.\ 2003.} \\
\multicolumn{7}{l}{$^k$Period confirmed by A.M. Levine and D. Lin, private communication.} \\
\multicolumn{7}{l}{$^l$``Q'' denotes quasi-persistent intervals (e.g., decades), rather than typical outburst.} \\
\end{tabular}
\end{table}

An inspection of Table~1 shows that 15 of the 20 X-ray sources have
values of $f(M)$ that require a compact object with a mass
$\gtrsim$3~\msun~.  This is a widely agreed limit for the maximum
stable mass of a neutron star in GR (e.g., Kalogera \& Baym 1996).
For the remaining five systems, some additional data are required to
make the case for a BH (Charles \& Coe 2006; McClintock \& Remillard
2006).  Historically, the best available evidence for the existence of
BHs is dynamical, and the evidence for these 20 systems is generally
very strong, with cautions for only two cases: LMC~X--1 and
XTE~J1859+226 (see McClintock \& Remillard 2006).  Thus, assuming that
GR is valid in the strong-field limit, we choose to refer to these
compact primaries as BHs, rather than as BH candidates.

\begin{figure}[ht]       
\centerline{\psfig{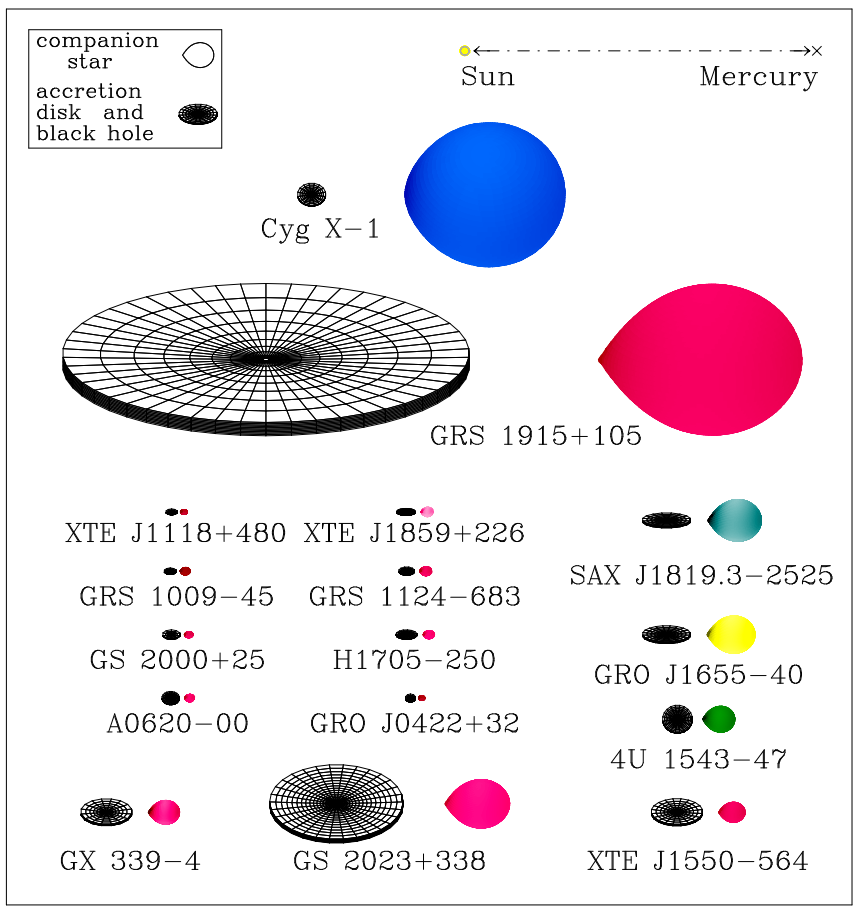}}
\caption{Scale drawings of 16 black-hole binaries in the Milky Way (courtesy of
J. Orosz).  The Sun--Mercury distance (0.4 AU) is shown at the top.
The estimated binary inclination is indicated by the tilt of the
accretion disk. The color of the companion star roughly indicates its
surface temperature.}
\label{fig:bhb}
\end{figure}
 
Figure~\ref{fig:bhb} is a schematic sketch of 16 Milky Way BHBs with
reasonably accurate dynamical data.  Their diversity is evident: there
are long-period systems containing hot and cool supergiants (Cyg X--1
and GRS~1915+105) and many compact systems containing K-dwarf
secondaries.  Considering all 20 BHBs (Table~1), only 3 are
persistently bright X-ray sources (Cyg X--1, LMC X--1 and LMC X--3).
The 17 transient sources include 2 that are unusual. GRS~1915+105
has remained bright for more than a decade since its first known
eruption in August 1992.  GX~339--4 undergoes frequent outbursts
followed by very faint states, but it has never been observed to fully
reach quiescence (Hynes et al.\ 2003).

The second half of Table~1 lists 20 X-ray binaries that lack radial
velocity data.  In fact, most of them even lack an optical
counterpart, and only three have known orbital periods.  Nevertheless,
they are considered black-hole candidates (BHCs) because they closely
resemble BHBs in their X-ray spectral and temporal behavior
(McClintock \& Remillard 2006; Tanaka \& Lewin 1995).  Some X-ray and
radio characteristics of the BHCs are given in Table 4.3 in McClintock
\& Remillard (2006); also given there is a subjective grade (A, B or
C) indicating the likelihood that a particular system does contain a
BH primary.  The seven A-grade BHCs are: A~1524--617, 4U~1630--47,
GRS~1739--278, 1E~1740.7--2942, H~1743--322, XTE J1748--288, and
GRS~1758--258.

\section{X-ray Observations of Black-Hole Binaries}

We first discuss X-ray light curves obtained by wide-angle X-ray
cameras that are used to discover X-ray novae and to monitor hundreds
of sources on a daily basis.  We then discuss timing and spectral
analyses of data obtained in pointed observations that reveal, in
detail, the properties of accreting BHs. Throughout this work we
feature \xte results derived from the huge and growing archive
of data amassed since 1996 by the All-Sky Monitor (ASM) and the
large-area Proportional Counter Array (PCA) detector (Swank 1998).

\subsection{X-ray Light Curves of Black-Hole Binaries in Outburst}

Nearly all BHBs are X-ray novae (see \S2 and Table~1) that are
discovered when they first go into outburst.  Their discovery and
subsequent daily monitoring are largely dependent on wide-field X-ray
cameras on orbiting satellites. The light curves of all 20 BHBs and
many BHCs (Table~1) can be found either in McClintock \& Remillard
(2006) or in a review paper on pre-RXTE X-ray novae by Chen et
al.\ (1997).  These researchers discuss the striking morphological
diversity among these light curves, which show broad distributions in
their timescales for rise and decay.

For X-ray outbursts that last between $\sim$ 20-days and many months,
the generally accepted cause of the outburst cycle is an instability
that arises in the accretion disk.  When the accretion rate from the
donor star is not sufficient to support continuous viscous flow to the
compact object, matter fills the outer disk until a critical surface
density is reached and an outburst is triggered. This model was
developed initially for dwarf novae (e.g., Smak 1971; Lasota 2001) and
extended to X-ray novae (e.g., Dubus et al.\ 2001).

This model predicts recurrent outbursts; indeed, half of the BHBs are
now known to recur on timescales of 1 to 60 years (Table~1).
Outbursts on much shorter or longer timescales do occur, but these are
not understood in terms of the disk instability model.  Sources such
as GRS~1915+105 and 4U~1755-338 exhibit ``on'' and ``off'' states that
can persist for $\gtrsim10$ years.  The behavior of the companion star
may play a role in causing these long-term changes in the accretion
rate.

After a decade of continuous operation, the ASM continues to scan most
of the celestial sphere several times per day (Levine et al.\ 1996).
It has discovered 8 BHB/BHC X-ray novae and an additional 15 recurrent
outbursts. Detailed X-ray light curves have been archived for each of
these sources, and complete outbursts for six such systems are shown
in \S5.

\subsection{X-ray Timing}

Our most important resource for examining the near-vicinity of a BH is
the rapid variations in X-ray intensity that are so often observed
(van der Klis 2005; McClintock \& Remillard 2006).  The analysis tool
commonly used for probing fast variability is the power-density
spectrum (PDS; e.g., Leahy et al.\ 1983).  Related techniques for
computing coherence and phase lag functions are reviewed by Vaughan \&
Nowak (1997). PDSs are interpreted with the presumption that the
source variations are a locally stationary process. More generalized
considerations of time series analyses are given by Scargle (1981),
while recent topics in non-linear processes are well described by
Gliozzi, Papadakis, \& R{\"a}th (2006).

The PDS is used extensively in this work.  The continuum power in the
PDS is of interest for both its shape and its integrated amplitude
(e.g., 0.1--10 Hz), which is usually expressed in units of rms
fluctuations scaled to the mean count rate.  PDSs of BHBs also exhibit
transient, discrete features known as quasi-periodic oscillations
(QPOs) that may range in frequency from 0.01 to 450 Hz.  QPOs are
generally modeled with Lorentzian profiles, and they are distinguished
from broad power peaks using a coherence parameter, $Q = \nu /
FWHM~\gtrsim~2$ (Nowak 2000; van der Klis 2005).  PDSs are frequently
computed for a number of energy intervals.  This is an important step
in linking oscillations to an individual component in the X-ray
spectrum.

\subsection{X-ray Spectra}

It has been known for decades that the energy spectra of BHBs often
exhibit a composite shape consisting of both a thermal and a
nonthermal component.  Furthermore, BHBs display transitions in which
one or the other of these components may dominate the X-ray luminosity
(see Tanaka \& Lewin 1995; McClintock \& Remillard 2006).  The thermal
component is well modeled by a multitemperature blackbody, which
originates in the inner accretion disk and often shows a
characteristic temperature near 1 keV (see \S7).  The nonthermal
component is usually modeled as a power law (PL).  It is characterized
by a photon index $\Gamma$, where the photon spectrum is $N(E) \propto
E^{-\Gamma}$.  The PL generally extends to much higher photon energies
($E$) than does the thermal component, and sometimes the PL suffers a
break or an exponential cutoff at high energy.

X-ray spectra of BHBs may also exhibit an Fe K$\alpha$ emission line
that is often relativistically broadened (\S8.2.3). In some BHBs,
particularly those with inclinations that allow us to view the disk
largely face-on, the spectral model requires the addition of a disk
reflection component (e.g., Done \& Nayakshin 2001).  In this case,
the X-ray PL is reflected by the accretion disk and produces a
spectral bump at roughly 10 to 30 keV.  Finally, high-resolution
grating spectra of BHBs sometimes reveal hot gas that is local to the
binary system (e.g., Lee et al 2002).  Such features may eventually
help us to interpret X-ray states, but at present there are too few
results to support any firm conclusions.

\section{Emission States of Black-Hole Binaries}

\subsection{Historical Notes on X-ray States}

The concept of X-ray states was born when Tananbaum et al.\ (1972)
observed a global spectral change in Cyg X--1 in which the soft X-ray
flux (2--6~keV) decreased by a factor of 4, the hard flux (10--20~keV)
increased by a factor of 2, and the radio counterpart turned on.
Thereafter, a similar X-ray transition was seen in A~0620--00 (Coe et
al.\ 1976) and in many other sources as well.  The soft state, which
was commonly described as $\sim$1 keV thermal emission, was usually
observed when the source was bright, thereby prompting the name
``high/soft state''. The hard state, with a typical photon index
$\Gamma \sim 1.7$, was generally seen when the source was faint, hence
the name ``low/hard state''.  In this state, the disk was either not
observed above 2 keV, or it appeared much cooler and withdrawn from
the BH.  An additional X-ray state of BHBs was identified in the {\it
Ginga} era (Miyamoto \& Kitamoto 1991; Miyamoto et al.\ 1993). It was
characterized by the appearance of several-Hz X-ray QPOs, a relatively
high luminosity (e.g., $> 0.1 L_{\rm Edd}$), and a spectrum comprised
of both a thermal component and a PL component that was steeper
($\Gamma \sim2.5$) than the hard PL.  This state was named the ``very
high'' state.

Rapid observational developments in the \xte era challenged the
prevailing views of X-ray states in BHBs. First, it was shown that the
soft state of Cyg X--1 is not consistent with a thermal interpretation
(Zhang et al.\ 1997b); instead, the spectrum is dominated by a steep
PL component ($\Gamma \sim 2.5$). Thus, Cyg X--1 is not a useful
prototype for the high/soft state that it helped to define. Secondly,
the spectra of BH transients near maximum luminosity were often found
to exhibit a steep PL spectrum, rather than a thermal spectrum
(McClintock \& Remillard 2006).  Thirdly, a number of different QPO
types were commonly observed over a wide range of luminosities (e.g.,
Morgan et al.\ 1997; Sobczak et al.\ 2000a; Homan et al.\ 2001).
These findings attracted great interest in the nature of the very high
state.

During this period, Gamma-ray observations ($\sim$40--500~keV) of
seven BHBs brought clarity to the distinction between the soft and
hard types of X-ray PL components (Grove et al.\ 1998; Tomsick et
al. 1999).  Sources in the low/hard state ($\Gamma \sim 1.7$) were
found to suffer an exponential cutoff near 100~keV, whereas sources
with soft X-ray spectra ($\Gamma \sim 2.5$) maintained a steep, strong
and unbroken PL component out to the sensitivity limit of the
gamma-ray detectors ($\sim 1$ MeV).

More recently, radio observations have cemented the association of the
low/hard state with the presence of a compact and quasi-steady radio
jet (see Fender 2006; McClintock \& Remillard 2006).  In brief, the
evidence for the association includes the following: ($a$) the
presence of compact jets in VLBI images of two BH sources (Dhawan,
Mirabel, \& Rodriguez 2000; Stirling et al.\ 2001); ($b$) correlated
X-ray and radio intensities and/or the presence of flat or inverted
radio spectra (e.g., Gallo et al.\ 2003), which allow the jet's
presence to be inferred even in the absence of VLBI images; ($c$) a
2\% linear radio polarization at nearly constant position angle
observed for GX~339--4 (Corbel et al.\ 2000); and ($d$) the
frequently-observed quenching of the persistent radio emission that
occurs when a BHB switches from the low/hard state to the high/soft
state (e.g., Fender et al.\ 1999).

\subsection{A Quantitative Three-State Description for Active 
Accretion}

\begin{figure}[ht]       
\centerline{\psfig{figure=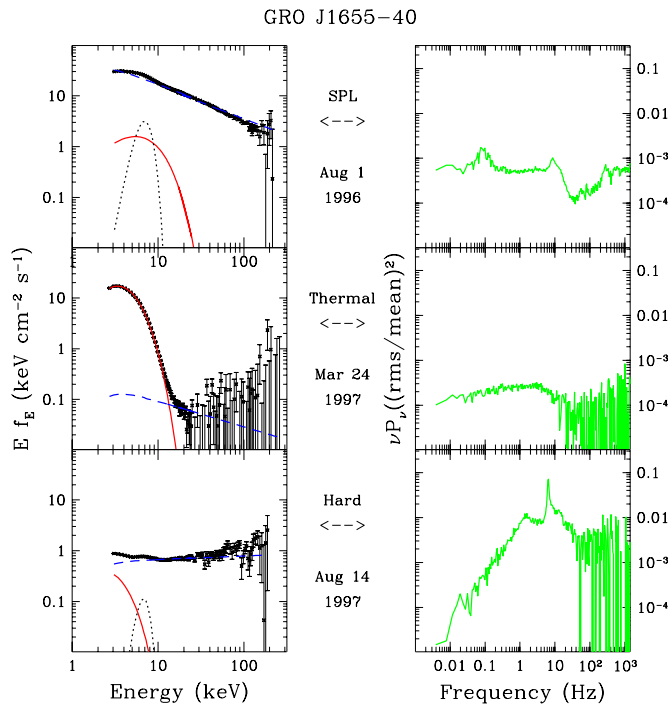,height=4in}}
\caption{Sample spectra of black-hole binary GRO~J1655--40
illustrating the three outburst states: steep power law, thermal, and
hard.  Each state is characterized by a pair of panels.  Left panels
show the spectral energy distribution decomposed into three
components: thermal ($red,~solid~line$), power-law ($blue,~dashed~line$),
and a relativistically broadened Fe~K$\alpha$ line ($black,~dotted~
line$).  Right panels show the PDSs plotted as log($\nu \times
P_{\nu}$) versus log$\nu$.}
\label{fig:xstates}
\end{figure}

In McClintock \& Remillard (2006), a new framework was used to define
X-ray states that built on the preceding developments and the very
extensive \xte data archive for BHBs.  In Figure~\ref{fig:xstates}, we
illustrate the character of each state by showing examples of PDSs and
energy spectra for the BHB GRO~J1655--40.  The relevance of X-ray
states fundamentally rests on the large differences in the energy
spectra and PDSs that can be seen in a comparison of any two 
states. For thorough discussions, many illustrative spectra, and
detailed references, see McClintock \& Remillard (2006).  Discussions
of physical models for these states are given in \S7.

The sharpest point of departure from the old description of states is
that luminosity is abandoned as a criterion for defining the
state of a source.  In defining states, McClintock \& Remillard
(2006) adopted the pragmatic and generic strategy of utilizing a
spectral model consisting of a multitemperature accretion disk and a
PL component (with a possibile break near 15 keV or an exponential
cutoff in the range 30--100 keV).  When required, an Fe emission line
or a reflection component was included.  The model also included
photoelectric absorption by neutral gas.  McClintock \& Remillard
(2006) used four parameters to define X-ray states: (1) the disk
fraction $f$, which is the ratio of the disk flux to the total flux
(both unabsorbed) at 2--20 keV; (2) the PL photon index ($\Gamma$) at
energies below any break or cutoff; (3) the rms power ($r$) in the PDS
integrated from 0.1--10 Hz, expressed as a fraction of the average
source count rate; and (4) the integrated rms amplitude ($a$) of any
QPO detected in the range 0.1--30 Hz. PDS criteria ($a$ and $r$)
utilize a broad energy range, e.g., the bandwidth of the \xte~PCA
instrument, which is effectively 2--30 keV.  Quantitative definitions
of the three states are given in Table~2.

\begin{table}
\caption{Outburst states of black holes: nomenclature and definitions}
\vspace*{0.15cm}
\small
\begin{tabular}{ll}
\toprule
{\bf New State Name}        &Definition of X-ray State$^a$ \\
(Old State Name)            & \\
\colrule
{\bf Thermal}               &Disk fraction $f^b > 75$\% \\
(High/Soft)                 &QPOs absent or very weak: $a_{\rm max}^c < 0.005$ \\
                            &Power continuum level $r^d < 0.075$$^e$ \\
                            & \\
{\bf Hard}                  &Disk fraction $f^b < 20$\% (i.e., Power-law fraction $>$ 80\%) \\
(Low/Hard)                  &$1.4^f < \Gamma < 2.1$ \\
                            &Power continuum level $r^d > 0.1$ \\
                            & \\
{\bf Steep Power Law (SPL)} &Presence of power-law component with $\Gamma > 2.4$ \\
(Very high)                 &Power continuum level $r^d < 0.15$ \\
                            &{\it Either} $f^b < 0.8$ and 0.1--30 Hz QPOs present with $a^c > 0.01$\\
                            &{\it or} disk fraction $f^b < 50$\% with no QPOs \\
\botrule
\multicolumn{2}{l}{$^a$2--20 keV band.} \\
\multicolumn{2}{l}{$^b$Fraction of the total 2--20 keV unabsorbed flux.} \\
\multicolumn{2}{l}{$^c$QPO amplitude (rms).} \\
\multicolumn{2}{l}{$^d$Total rms power integrated over 0.1--10 Hz.} \\
\multicolumn{2}{l}{$^e$Formerly 0.06 in McClintock \& Remillard (2006).} \\
\multicolumn{2}{l}{$^f$Formerly 1.5 in McClintock \& Remillard (2006).} \\
\end{tabular}
\end{table}

In the thermal state (formerly high/soft state, and ``thermal
dominant'' state in McClintock \& Remillard 2006) the flux is
dominated by the heat radiation from the inner accretion disk, the
integrated power continuum is faint, and QPOs are absent or very weak
(see Table~2).  There is usually a second, nonthermal component in the
spectrum, but its contribution is limited to $< 25$\% of the flux at
2-20 keV.  The state is illustrated in the middle row of panels in
Figure~\ref{fig:xstates}.  The spectral deconvolution shows that the
thermal component ($red~line$) is much stronger than the PL component
($blue~dashed~line$) for $E \lesssim 10$ keV.  The PDS ($right~panel$),
which is plotted in terms of log($\nu \times P_{\nu}$) versus
log$\nu$, appears featureless. Similar displays of paired energy
spectra and PDSs for nine other BHBs/BHCs in the thermal state are
shown in McClintock \& Remillard (2006).

The hard state (formerly low/hard state) is characterized by a hard PL
component ($\Gamma \sim 1.7$) that contributes $\geq 80$\% of the
2--20 keV flux (Table~2).  The power continuum is bright with $r >
0.1$ and QPOs may be either present or absent.  A hard state
osbervation of GRO~J1655--40 is shown in the bottom row of panels in
Figure~\ref{fig:xstates}.  The accretion disk appears to be faint and
cool compared to the thermal state.  As noted previously (\S4.1), the
hard state is associated with the presence of a quasi-steady radio
jet, and clear correlations between the radio and X-ray intensities
are observed.

The hallmark of the steep power law (SPL) state (formerly the very
high state) is a strong PL component with $\Gamma \sim 2.5$.  In some
sources, this PL has been detected without a break to energies of
$\sim1$~MeV or higher (\S4.1).  This state is also characterized by
the presence of a sizable thermal component and the frequent presence
of X-ray QPOs (see Table~2).  An example of the SPL state is shown in
the top row of panels in Figure~\ref{fig:xstates}, and many additional
illustrations are displayed in McClintock \& Remillard (2006).  There
are similarities between the SPL state and the thermal state; both
show a thermal component and a steep PL component.  However in the
thermal state the PL is faint and has a more variable photon index,
while the SPL state is plainly distinguished by its powerful PL
component and the commonly-occurring QPOs.  The SPL state tends to
dominate BHB spectra as the luminosity approaches the Eddington limit,
and it is this state that is associated with high-frequency QPOs
(\S6.2; McClintock \& Remillard 2006).

Intermediate states and state transitions are another important aspect
of BHB studies. The three states defined by McClintock \& Remillard
(2006) attempt to specify spectral and timing conditions
that are quasi-stable and that appear to have distinct physical
origins.  There are gaps in the parameter ranges used to define these
states (Table~2), and this gives rise to intermediate states. The
hybrid of the hard and SPL states is particularly interesting for its
correlations with radio properties (\S4.3 and \S5) and also with disk
properties (\S6.1).

\subsection{The Unified Model for Radio Jets}

Many researchers investigate the spectral evolution of BHBs using a
hardness-intensity diagram (HID), which is a plot of X-ray intensity
versus a ``hardness ratio'' ($HR$), i.e., the ratio of detector counts
in two energy bands (e.g., van der Klis 2005; Homan et al.\ 2001;
Belloni 2004).  This diagram is widely used in tracking the behavior
of accreting neutron stars.  Compared to the spectral-fitting approach
described above, the HID approach has the advantage that it is model
independent and the disadvantage that it is difficult to relate the
results to physical quantities. Interpretations of variations in the
HID depend on the particular energy bands chosen to define $HR$ in a
given study.  If both bands are above $\sim$5 keV, then the $HR$ value
effectively tracks the slope of the PL component (i.e., lower $HR$ is a
steeper PL). Softer energy bands admit a mixture of thermal and
nonthermal components, and interpretations are then more complicated.

\begin{figure}[ht]       
\centerline{\psfig{figure=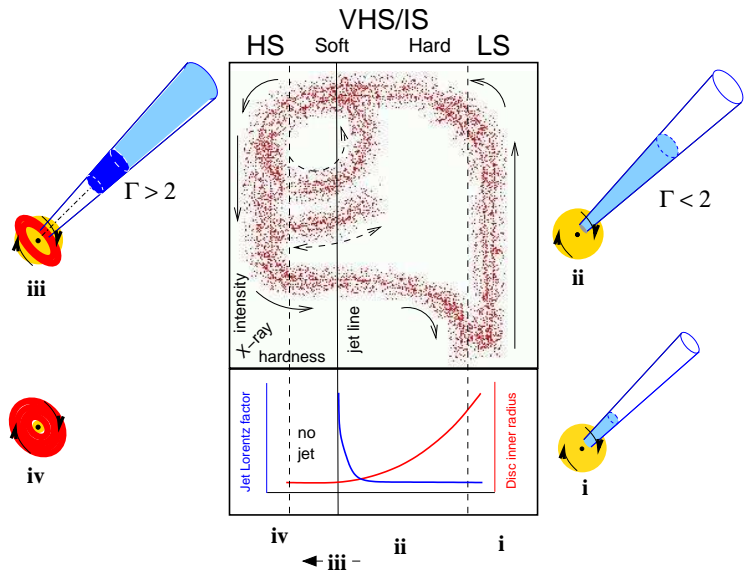,height=3in}}
\caption{A schematic representation of the model for disk-jet coupling
in black-hole binaries from Fender et al. (2004).  The top panel shows
evolutionary tracks in a HID, which is a plot of X-ray intensity
versus X-ray hardness.  (These quantities increase upward and to the
right, respectively.)  The bottom panel gives a qualitative impression
of how the jet's bulk Lorentz factor ($blue~curve$) and the inner disk
radius ($red~curve$) vary with X-ray hardness.  The X-ray states are
labeled at the top in the old nomenclature (i.e., HS = high/soft
state; VHS/IS = very high and intermediate states; and LS = low/hard
state.)}
\label{fig:unifjet}
\end{figure}

The HID is also used in illustrating the ``unified model for radio
jets'' proposed by Fender et al.\ (2004). Figure~\ref{fig:unifjet}
shows their schematic for the relationships between jets and X-ray
states, where the state of an observation is distinguished simply by
the value of $HR$.  The figure shows qualitatively how the jet Lorentz
factor ($lower~panel$) and the morphology of the jet ($sketches~i-iv$)
evolve with changes in the X-ray state.  Tracks for state transitions
of an X-ray source in the HID are also shown ($top~panel$).  The solid
vertical line running through both panels in Figure 3 is the ``jet
line.''  To the right of the jet line the X-ray spectrum is relatively
hard and a steady radio jet is present, and to the left the spectrum
is soft and the jet is quenched.  The jet line also marks an
instability strip where violent ejections of matter may occur (see
$sketch~iii$), as indicated by the spike in the Lorentz factor
($lower~panel$).  The tracks for state evolution in the HID are
influenced by observations of GX~339--4 (Belloni et al. 2005).  In the
Belloni et al. study, $HR$ was defined as the ratio of source counts
at 6.3--10.5 keV to the counts at 3.8--7.3 keV.

The vertical source track on the right side of Figure 3 corresponds to
the low/hard or hard state with $HR > 0.8$ for GX~339--4 (see Belloni
it al. 2005).  The vertical track on the far left (and most
observations of GX~339--4 with $HR < 0.2$) are in the high/soft or
thermal state.  Thus, the HID state classifications and the McClintock
\& Remillard (2006) classifications agree very well in these two
regimes.  However, for the intermediate values of $HR$, which fall
between the dashed lines, the states are described differently.  The
states in this region are further divided by Belloni et al. (2005)
into ``soft intermediate'' and ``hard intermediate'' states, based on
the $HR$ values and the properties of the PDS power continuum.  This
completes a description of the four X-ray states defined in the
unified jet model.  Further comparisons of the HID state
classifications and the McClintock \& Remillard (2006) classifications
are given in \S5.

The unified model for X-ray states and radio jets provides
opportunities to study the disk-jet coupling explicitly, and the HID
format is very easy to apply to observations. On the other hand, the
state definitions of McClintock \& Remillard (2006) are more
quantitative, and they provide spectral information that is more
directly applicable to physical models.  In \S5, we present outburst
data for selected BHB/BHC systems using both depictions for X-ray
states.

\subsection{Quiescent State}

The quiescent state corresponds to luminosities that are three or
more orders of magnitude below the levels of the active states described
above.  The typical BHB with transient outbursts spends most of its time
in a quiescent state that is characterized by an extraordinarily
faint luminosity ($L_x = 10^{30.5}-10^{33.5}$ \ergsec) and a
spectrum that is distinctly nonthermal and hard ($\Gamma =
1.5-2.1$). The quiescent state is particularly important in two ways:
($a$) It enables firm dynamical measurements to be made because the
optical spectrum of the secondary star becomes prominent and is
negligibly affected by X-ray heating (van Paradijs \& McClintock 1995);
and ($b$) its inefficient radiation mechanism underpins a strong argument
for the event horizon (\S8.1).

For a thorough review of the X-ray properties of this state and
discussions of physical models, see McClintock \& Remillard (2006).
As commonly remarked, it is possible that the hard and quiescent
states represent a single mechanism that operates over several orders
of magnitude in X-ray luminosity. However, this question remains
controversial.  Corbel, Tomsick \& Kaaret (2006) have recently derived
precise spectral parameters for XTE~J1550-564. They conclude that the
quiescent spectrum of this source (and a few others) is softer than
the spectrum in the much more luminous hard state.

\section{X-ray Overview of State Evolution and Energetics}

\subsection{Overview Plots for Six Individual Sources}

In \S4, we described two ways of defining X-ray states: a quantitative
method based on generic X-ray spectral modeling and PDS analysis
(McClintock \& Remillard 2006), and another based on radio properties,
X-ray PDSs, and HIDs.  Here we synthesize the two approaches to
provide the reader with a comprehensive picture of the behavior of an
accreting BH.

\begin{figure}[ht!]            
\centerline{\psfig{figure=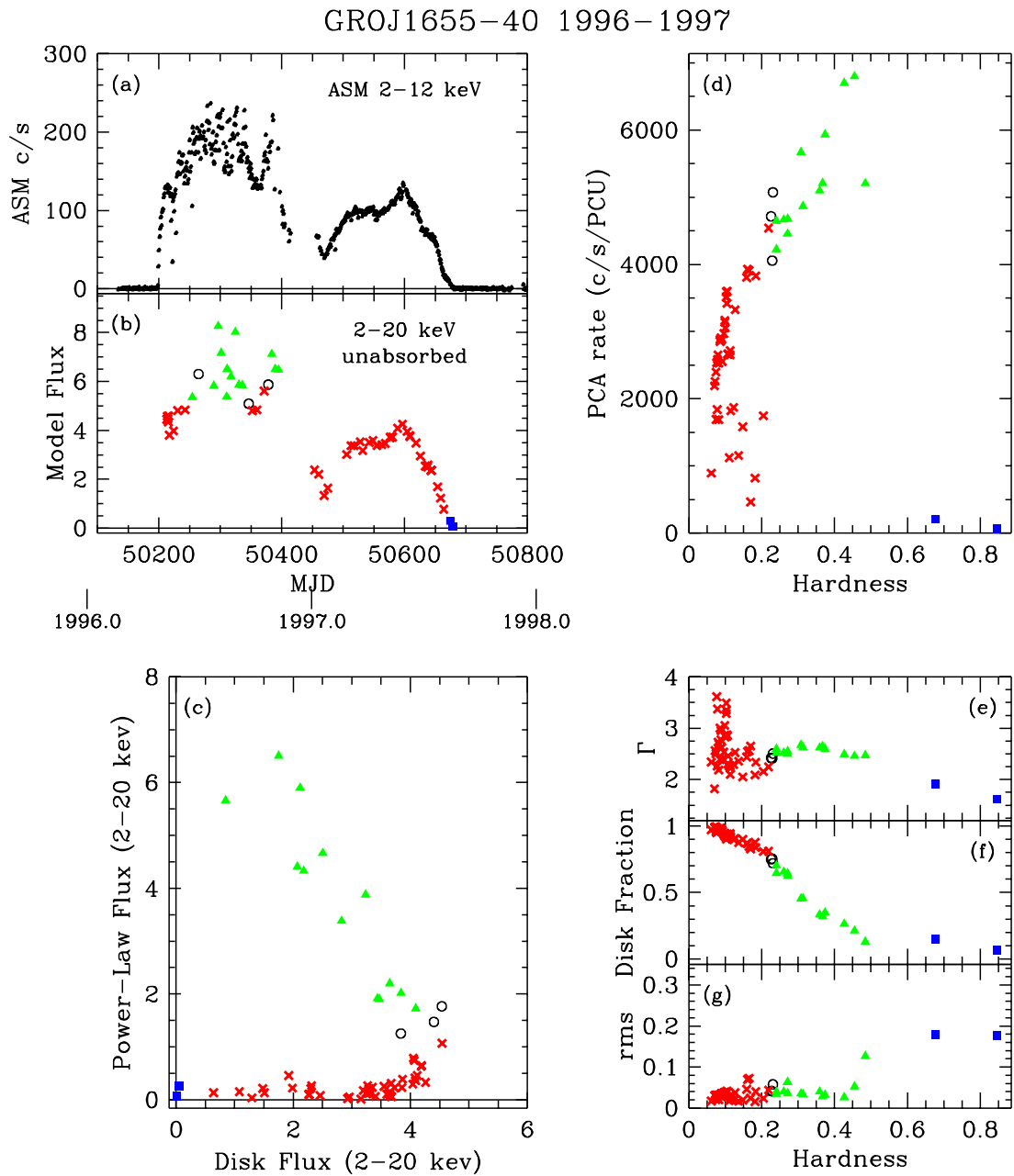,height=7in}}
\caption{X-ray overview of GRO~J1655--40 during its 1996--1997
outburst.  The ASM light curve is shown in panel $a$. All other data
are derived from \xte pointed observations binned in 62 intervals. The
symbol color and type denote the X-ray stat: thermal ($red~x$), steep
power law (SPL) ($green~triangle$), hard ($blue~square$), and any
intermediate type ($black~circle$). This outburst shows simple
patterns of evolution that favor the thermal and SPL states.}
\label{fig:xo1655}
\end{figure}

\begin{figure}[ht!]           
\centerline{\psfig{figure=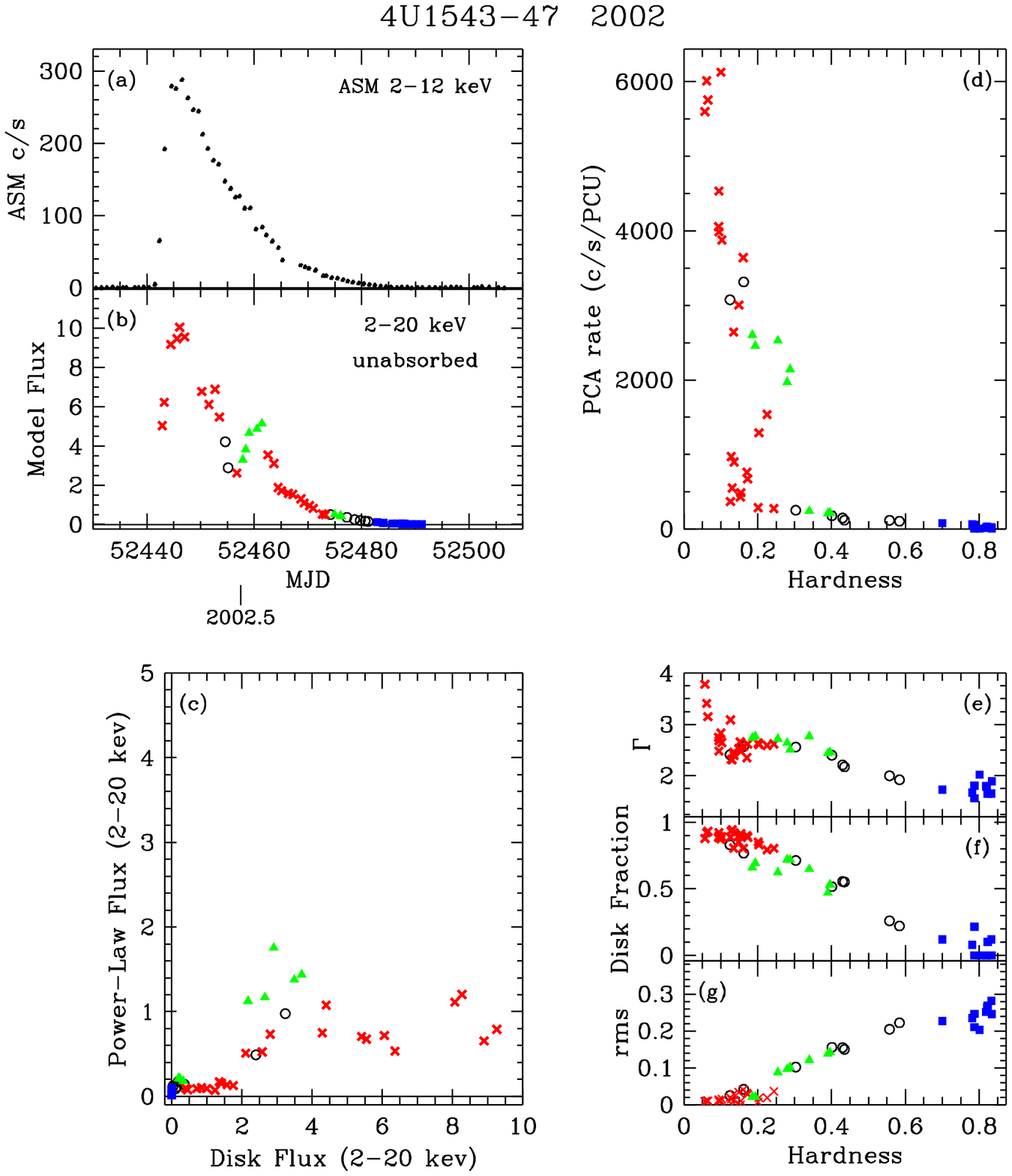,height=7in}}
\caption{X-ray overview of 4U~1543--47 during its 2002 outburst.
The presentation format and the state-coded plotting symbols follow the
conventions of Figure~\ref{fig:xo1655}.  The \xte pointed observations
(49 time intervals) show relatively simple patterns of state evolution,
and the thermal state is prevalent when the source is bright.}
\label{fig:xo1543}
\end{figure}

\begin{figure}[ht!]            
\centerline{\psfig{figure=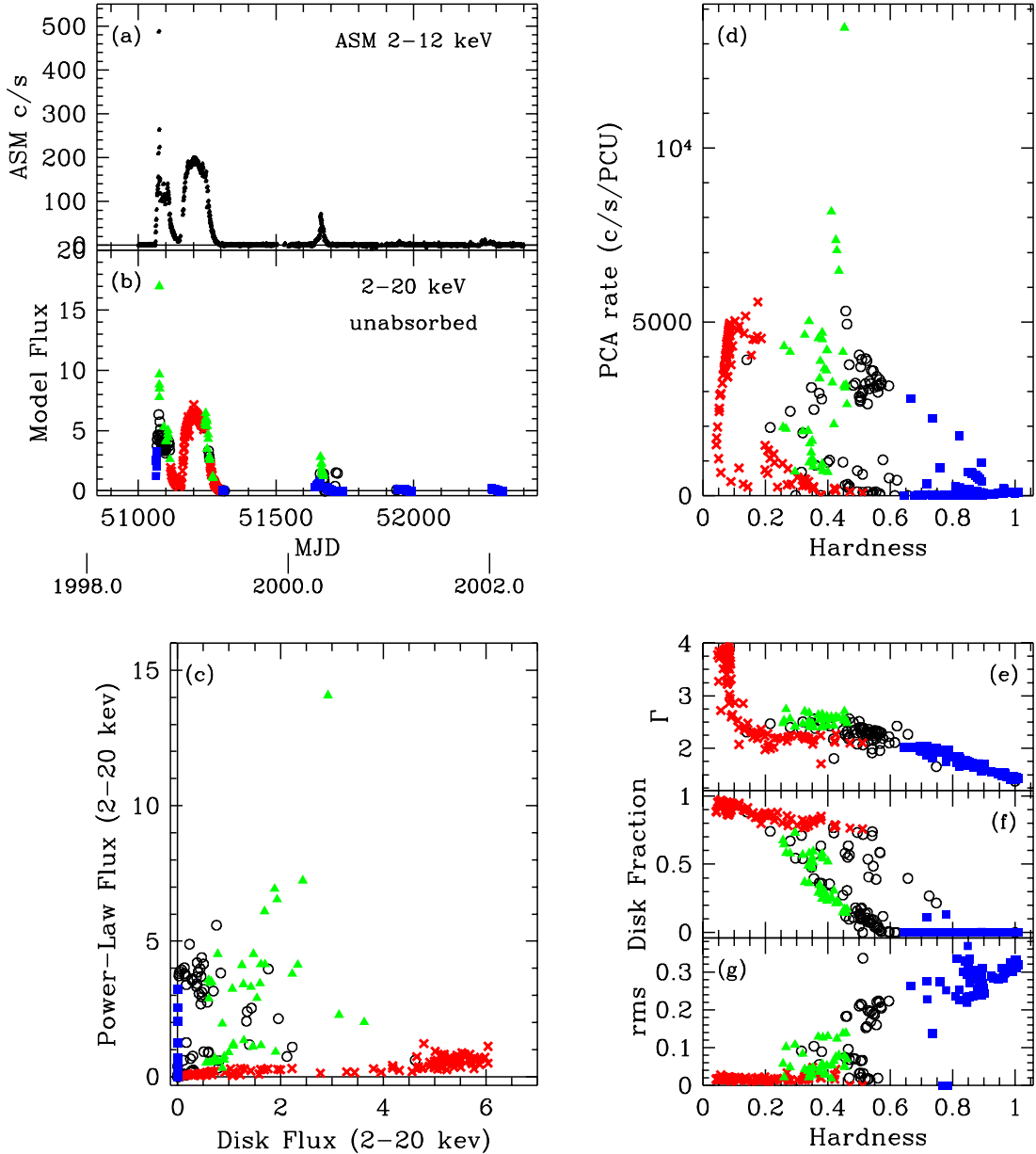,height=7in}}
\caption{X-ray overview (309 time intervals) of XTE~J1550--564
that includes a series of four outbursts with decreasing maxima. The
presentation format and the state-coded plotting symbols follow the
conventions of Figure~\ref{fig:xo1655}.  The two brighter outbursts
(1998--1999 and 2000) show great complexity in the temporal evolution
of states and the energy division between thermal and nonthermal
components.  In contrast, the subsequent pair of faint outbursts were
confined to the hard state.}
\label{fig:xo1550}
\end{figure}

\begin{figure}[ht!]          
\centerline{\psfig{figure=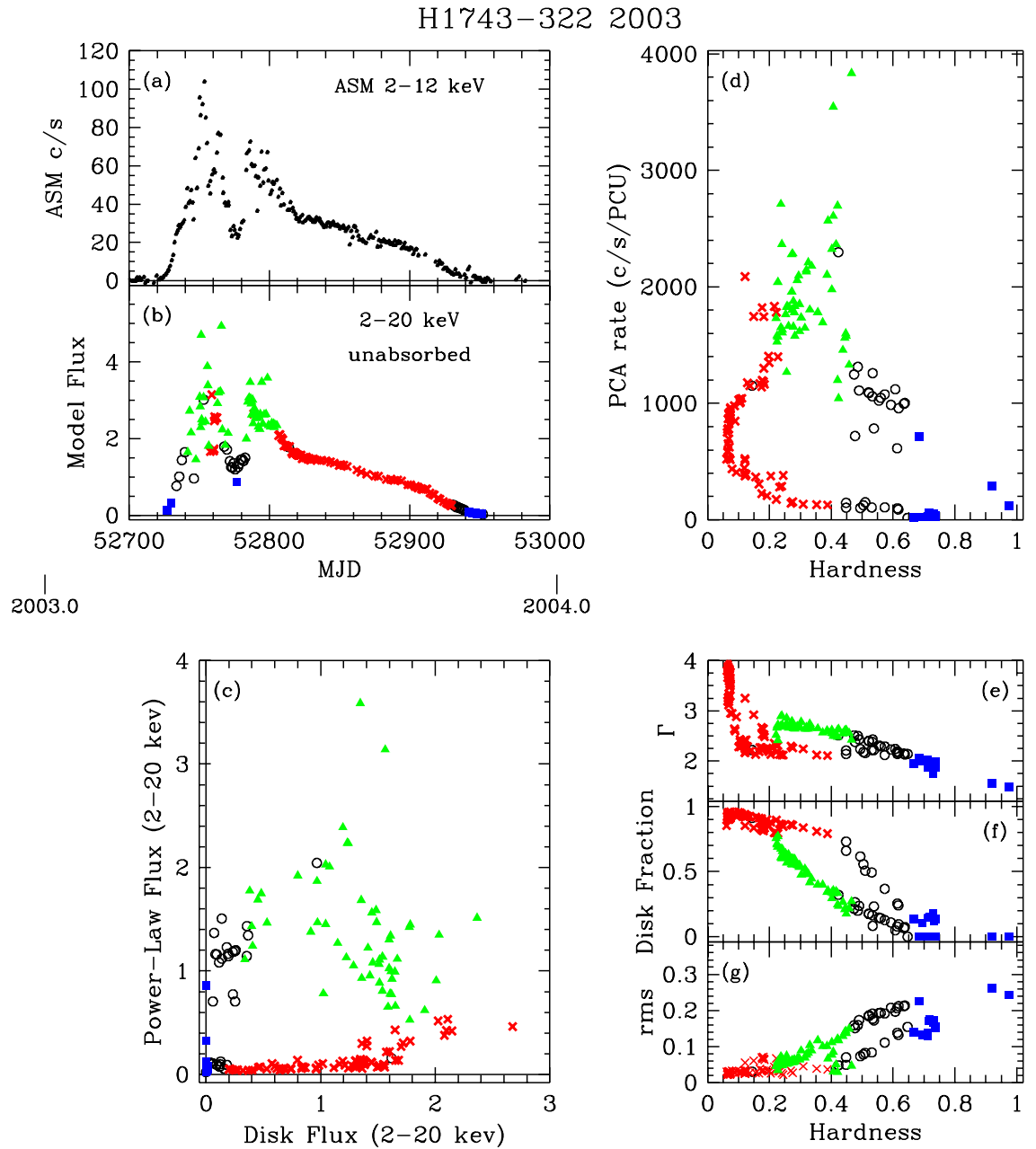,height=7in}}
\caption{X-ray overview of H~1743--322 during its 2003
outburst.  The presentation format and the state-coded plotting
symbols follow the conventions of Figure~\ref{fig:xo1655}.  The \xte
pointed observations (170 time intervals) show complex state evolution with
similarities to XTE~J1550--564.}
\label{fig:xoh1743}
\end{figure}

\begin{figure}[ht!]           
\centerline{\psfig{figure=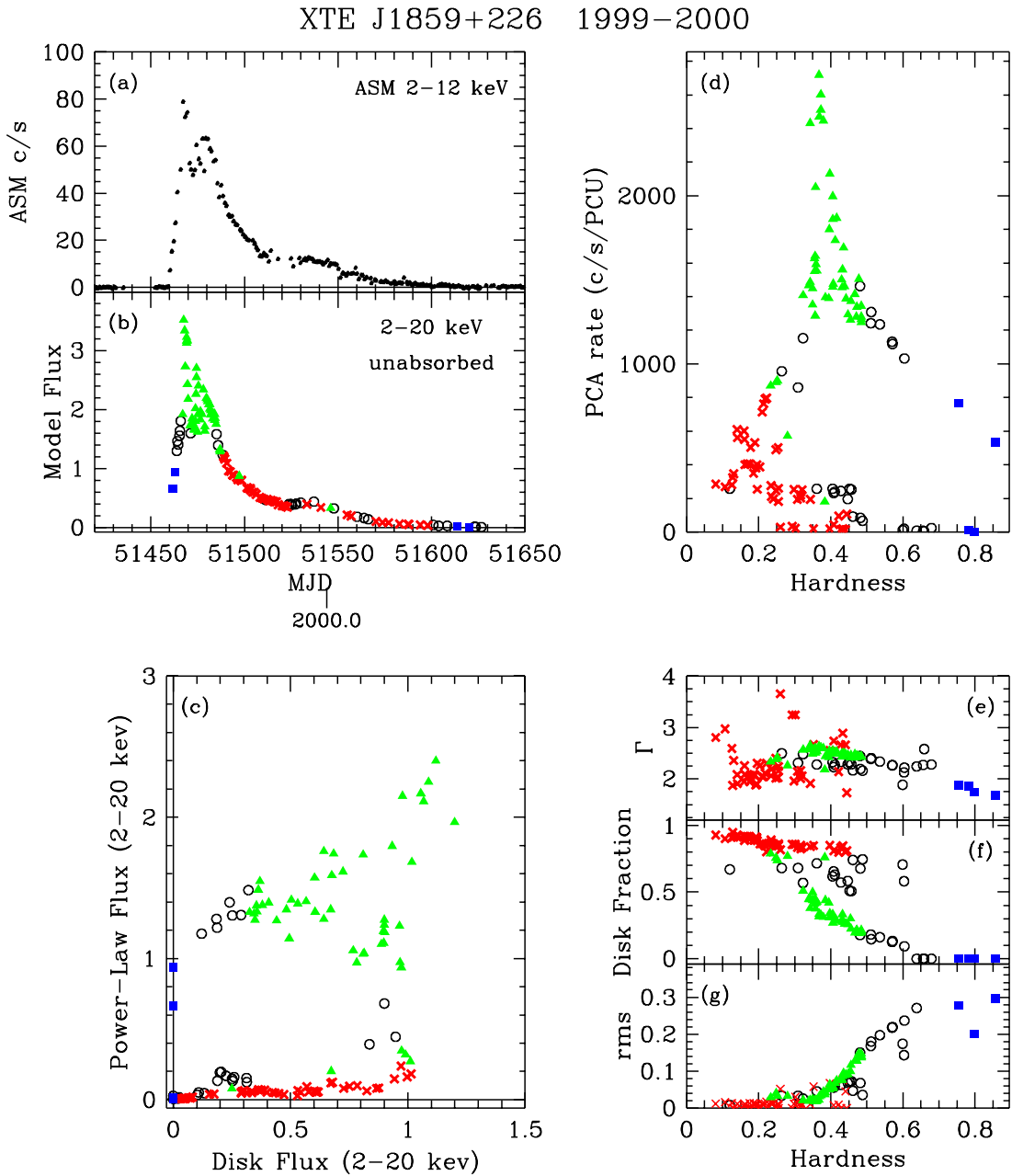,height=7in}}
\caption{X-ray overview of XTE~J1859+226 during its 1999--2000
outburst.  The presentation format and the state-coded plotting
symbols follow the conventions of Figure~\ref{fig:xo1655}.  The \xte
pointed observations (130 time intervals) show complex behavior very
similar to that seen for H~1743--322. In panels $b$, $c$, and $d$ it
is apparent that the steep power law state is prevalent when the
source is bright.}
\label{fig:xo1859}
\end{figure}

\begin{figure}[ht!]          
\centerline{\psfig{figure=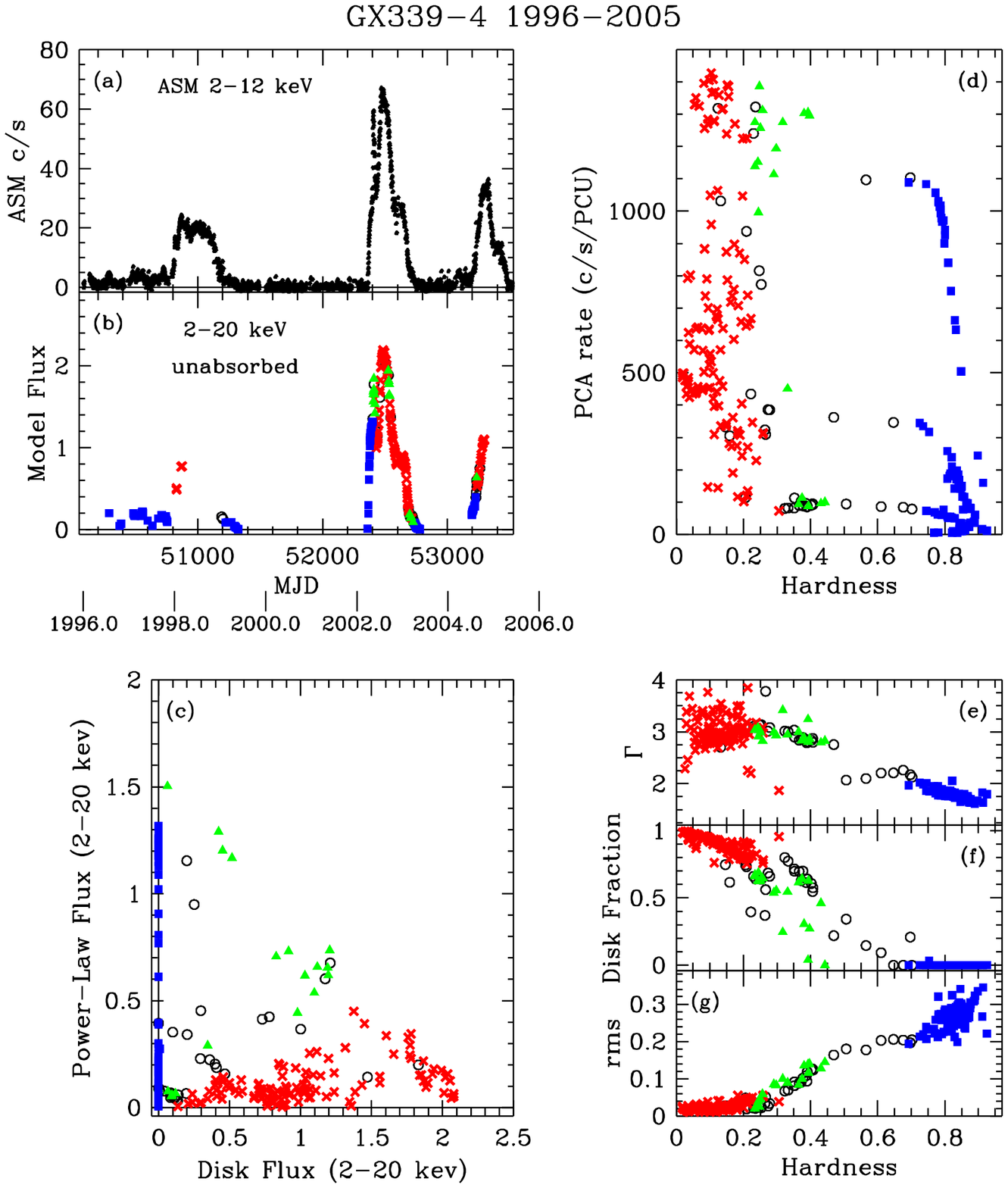,height=7in}}
\caption{X-ray overview of GX~339--4 (274 time intervals). This source
exhibits frequent outbursts and long intervals in the hard state at
low luminosity.  The presentation format and the state-coded plotting
symbols follow the conventions of Figure~\ref{fig:xo1655}.  The state
evolution shown here was used as the prototype for making the
connection between the inner disk and a steady radio jet (compare
panel $d$ and Figure~3).}
\label{fig:xogx339}
\end{figure}

In Figures~\ref{fig:xo1655}--\ref{fig:xogx339} respectively, we show
detailed overviews of the behavior of six X-ray novae: five BHBs and
one BHC (H~1743--322).  Each figure contains seven panels that review
the data for a single source.  The overview plots can be used in two
ways. One can focus on the figure (7 panels) for a particular BHB and
examine for each color-coded state many aspects of the behavior of the
source.  The key science questions addressed by each panel are:
($a,b$) How do states and luminosity vary with time?; ($c$) How does
the radiation energy divide between thermal and nonthermal components
(2-20 keV)?; ($d$) How do the states of McClintock and Remillard
(2006) relate to the states of Fender, Belloni, and Gallo (2004) which
are presented in a HID?; and ($e,f,g$) How do three key X-ray
properties -- PL index, disk fraction, and rms power -- vary as a
function of either the hardness ratio or the X-ray state?  On the
other hand, one can choose a particular panel and compare the behavior
of the six sources to draw general conclusions about common behavior
patterns in BHBs as well as their differences.  Such conclusions are
discussed at the end of this Section and in \S 5.2.

The following comments pertain to all six figures.  As an inspection
of the various ASM light curves ($panel~a$) shows, the data cover
outbursts observed with \xte during the time interval
1996--2004.  The remaining panels ($b-g$) show results derived from
\xte~pointed observations. These latter panels display several
different kinds of data, but in every panel there is a common use of
symbol type and color to denote the state of the source (McClintock \&
Remillard 2006 definitions): thermal ($red~x$), hard ($blue~
square$), SPL ($green~triangle$), and any intermediate type ($black~
circle$).

Panel $b$, which mimics the ASM light curve, shows the 2--20 keV X-ray
flux derived from the fitted spectral model (see \S3.3).  Panel $c$,
which is also based on the fitting results, shows how the energy is
divided between the thermal (accretion disk) component and the
nonthermal (PL) component; we refer to this plot as the
energy-division diagram.  Panel $d$, which is based on raw count
rates, shows how the states are distributed in the hardness intensity
diagram (HID).  We use this panel to examine how the McClintock \&
Remillard (2006) states are distributed in the HID and how they
correspond to the states that are defined by the unified jet model
(\S4.3).  Finally, in panels $e-g$ we plot spectral hardness on the
x-axis versus three of the parameters that are used to define BH
states, namely, the PL index $\Gamma$, the disk fraction, and the
integrated rms power in the PDS (0.1--10 Hz).  The HIDs use the
normalization scheme and hard color definition ($HR$ = 8.6--18.0 keV
counts / 5.0--8.6 keV counts) given by Muno et al. (2002), for which
the Crab Nebula yields 2500 c s$^{-1}$ PCU$^{-1}$ and $HR = 0.68$.
The \xte~pointed observations are selected to have a minimum exposure
Rtime of 500 s and a minimum source flux of $10^{-10}$ erg cm$^{-2}$
s$^{-1}$ (i.e., 4 mCrab) at 2--20 keV.

We begin our discussion of individual sources with GRO~J1655--40, the
BHB used in \S 4.2 to illustrate the BH states (McClintock \&
Remillard 2006). In Figure~\ref{fig:xo1655} we show the overview for
the 1996--1997 outburst.  The data are mostly derived from
publications (Sobczak et al.\ 1999; Remillard et al.\ 1999; 2002a),
with some supplementary results added for completeness.  GRO~J1655--40
shows an orderly and monotonic evolution of states along the arcs
displayed in the energy-division diagram ($panel~c$) and the HID
($panel~d$).  In the HID, the McClintock \& Remillard (2006) states
are cleanly sorted along the hardness axis.  Furthermore, panels
$e-g$ show clear correlations between the key state parameters and
the hardness ratio.  Quite similar behavior is seen in the X-ray
overview for the next BHB, 4U~1543--47 (Figure~\ref{fig:xo1543}). Here
we have used the results derived by Park et al (2004), while adding
results from similar analyses to extend the coverage to the end of the
outburst. As these two overviews show (i.e., Figures 5 and 6), both
sources favor the thermal state, and they only enter the hard state at
low luminosity.
 
Figure~\ref{fig:xo1550} exhibits four outbursts of XTE~J1550--564 with
successively shorter durations and decreasing maxima ($panels~a$ and
$b$).  The results shown in panels $c-g$ are dominated by the first
outburst (1998--1999), which provides 202 of 309 state assignments and
most of the points at high flux levels.  Several authors have noted
the complex behavior of this source (e.g., Sobczak et al.\ 2000b;
Homan et al.\ 2001).  The energy-division diagram ($panel~c$) and the
HID ($panel~d$) display many branches.  The thermal branch
($red~x~symbols$) in the energy diagram covers a wide range of
luminosity.  Note that there are examples of both high luminosity
hard-state observations and low luminosity SPL-state observations. One
must conclude that luminosity does not drive a simple progression of
X-ray states (as implied by the old state names: low/hard, high/soft
and very high).  Thus, the X-ray state must depend on some unknown and
important variable(s) in addition to the BH mass and the mass
accretion rate (Homan et al.\ 2001).

Similar patterns of behavior are shown in the next two examples.  The
overview for H~1743--322 (Figure~\ref{fig:xoh1743}, panels $b-g$)
shows many state transitions ($panel~b$), and the complex tracks in
panels $c$ and $d$ are reminiscent of XTE~J1550--564
(Figure~\ref{fig:xo1550}).  The hysteresis in transitions to and from
the hard state are especially evident in the HID ($panel~d$; see also
Maccarone \& Coppi 2003). The overview for XTE~J1859+226
(Figure~\ref{fig:xo1859}), is similar to that of H~1743--322. All of
the brightest observations are found in the SPL state, and the
vertical track in the HID ($green~triangles$ in
Figure~\ref{fig:xo1859}$d$) is a consequence of constant values of
$\Gamma$ in the SPL state. These complex results reinforce the need to
probe deeply in attempting to identify the underlying variables that
govern the emission of radiation from BHBs.
 
Finally, we show the overview of GX~339--4 in
Figure~\ref{fig:xogx339}. This source is known to produce frequent
X-ray outbursts and to remain for long intervals in the hard state, as
shown in panel $b$.  It is therefore a frequent target for radio
studies (e.g., Fender et al.\ 1999; Corbel et al.\ 2000; Gallo et al.\
2003; Belloni et al.\ 2005).  There is an apparent similarity between
the HID for GX~339--4 ($panel~d$) and to the schematic sketch in
Figure~\ref{fig:unifjet} that illustrates the unified model for jets.
The HID shows four branches for the hard state, one of which peaks at
75\% of the maximum count rate in the thermal and SPL states.  Despite
the wide range in luminosity for each of the three states, the plots
of the key parameters ($panels~e-g$) appear orderly and well correlated
with the hardness ratio.  Thus, the behavior of GX~339--4 supports the
theme that recurs throughout this work, namely, that X-ray states are
important, although they are plainly not a simple function of source
luminosity.
 
The six X-ray overviews presented in Figures 4--9 show that BHB
outbursts can be very complex. They typically begin and end in the
hard state, but between those times there is common disorder in the
temporal evolution of luminosity ($b$ panels), and in the division of
radiation energy between thermal and nonthermal states ($c$ panels).
On the other hand, there are clear correlations involving the key
spectral and timing properties, examined versus either $HR$ value or
the color-coded state symbols ($e$, $f$, and $g$ panels).  These
results, which are robust for a variety of BHBs, including sources
displaying multiple outbursts, confirm the prevailing wisdom that
complex BHB behavior can be productively organized and studied within
the framework of X-ray states.  Further discussions of the overview
figures are continued in the following section.

\subsection{Conclusions from Overviews of X-ray States}

Several conclusions can be drawn from the set of overview plots shown
in \S5. First, when the state assignments of McClintock \& Remillard
(2006) are examined in the HID format (i.e., sorting symbol color/type
vs. hardness in the $d$ panels of Figures 4--9), it is apparent that
these states and the designs of the unified-jet model overlap
significantly.  For example, for the particular $HR$ defined in \S5.1,
observations with $HR < 0.2$ would be assigned to the high/soft state
in the unified-jet model, and nearly all of these same observations
are here assigned to the thermal state of McClintock \& Remillard
(2006).  Similarly, observations with $HR > 0.65$ correspond to the
hard state in both prescriptions.  For the intermediate states in the
HID, there is a divergence between the two approaches, and this is
especially obvious for sources that show complex behavior, e.g.,
XTE~J1550--564.  For this source, the McClintock \& Remillard (2006)
states are strongly disordered in the interval $0.20 < HR < 0.65$ (see
Figure~\ref{fig:xo1550} $panel~d$).

We have emphasized that source luminosity is not a criterion used for
identifying X-ray states in either prescription (\S4.2 and \S4.3).
This is evident in Figures 4--9, because there are lines of constant
flux that intercept several different McClintock \& Remillard (2006)
states ($b$ panels), and because there are lines of constant intensity that
intercept observations with very different $HR$ values ($d$ panels).
However, it is clear that there is still some degree of correlation
between X-ray states and source brightness. For example, all of the
sources show transitions to the hard state when the source becomes
faint during its decay phase (see the $b$ and $d$ panels in Figures
4--9). At the opposite extreme, the highest luminosity observations
for four of the six sources occur in the SPL state (Figures 4 \&
6--8).

The energy-division diagrams (2--20 keV; $c$ panels of Figures 4--9)
routinely show vertical tracks for the hard state ($blue~squares$),
whereas horizontal tracks with gentle curvature are seen for the
thermal state ($red~x$ symbols).  These tracks indicate a free flow of
energy into the hard PL spectrum during the hard state and into the
accretion disk during the thermal state. This point was first made by
Muno et al.\ (1999) in describing the behavior of GRS~1915+105.  The
curvature of the thermal tracks ($c$ panels, Figures 4--9) can be
interpreted as being due to increased Comptonization with increasing
thermal luminosity.  Tracing smooth lines through these tracks would
constrain Comptonization to a maximum of $< 20$\% of the flux (2--20
keV) at the peak of the thermal state.  The energy-division tracks for
the SPL state ($green~triangles$) are not well defined, and the tracks
of different sources do not resemble each other. The energy-division
diagrams are more effective than the HIDs in delineating intrinsic
differences between the SPL and thermal states.

Comparisons between the SPL and thermal states in terms of the photon
index ($\Gamma$ ; $e$ panels of Figures 4--9) are of interest, as
these states are sometimes presumed to share a common PL mechanism
that is relatively muted during the thermal state. Generally, we find
broad distributions in $\Gamma$ during the thermal state, and in some
cases the thermal tracks are shifted from the SPL tracks (e.g.,
Figures 6$e$, 7$e$, 8$e$). We find no compelling evidence that the PL
mechanisms for the thermal and SPL states are the same. 

In the following section, we consider one additional observational
topic -- X-ray QPOs -- before discussing the ongoing efforts to relate
X-ray states to physical models for BH accretion.

\section{X-ray Quasi-Periodic Oscillations}

X-ray QPOs are specialized and extraordinarily important avenues for
the study of accreting BHs. They are transient phenomena associated
with the nonthermal states and state transitions.  For definitions of
QPOs and analysis techniques, see van der Klis (2005) and some details
in \S 3.2.

QPOs play an essential role in several key science areas, such as
probing regions of strong field (\S6.2 \& \S8.2.4) and defining the
physical processes that distinguish X-ray states.  Thus far in this
review, however, QPOs have been considered only tangentially (e.g.,
\S4.2).  In this section we focus on the significance of QPOs, their
subtypes, and spectral/temporal correlation studies that involve the
QPO frequency.  Following the literature, we divide the discussions of
QPOs into low-frequency and high-frequency groups.  In doing so, we
disregard the additional, infrequent appearances of very low frequency
QPOs (e.g., the QPO below 0.1 Hz in the upper-right panel of
Figure~\ref{fig:xstates}) that are not understood.  Physical models
for QPOs are briefly discussed in \S7, and the importance of
high-frequency QPOs for probing strong gravity is highlighted in
\S8.2.4.

\subsection{Low-Frequency Quasi-Periodic Oscillations}

Low-frequency QPOs (LFQPOs; roughly 0.1--30 Hz)) have been detected on
one or more occasions for 14 of the 18 BHBs considered in Table 4.2 of
McClintock \& Remillard (2006).  They are important for several
reasons.  LFQPOs can have high amplitude (integrated rms/mean values
of $a > 0.15)$ and high coherence (often $Q > 10$), and their
frequencies and amplitudes are generally correlated with the spectral
parameters for both the thermal and PL components (e.g., Muno et al.\
1999; Sobczak et al.\ 2000a; Revnivtsev et al.\ 2000; Vignarca et al.\
2003). With the exception of Cyg X--1, QPOs
generally appear whenever the SPL contributes more than 20\% of the
flux at 2--20 keV (Sobczak et al.\ 2000a), which is one component of
the definition of the SPL state (Table~2; McClintock \& Remillard
2006).

LFQPOs can vary in frequency on timescales as short as one minute
(e.g., Morgan et al.\ 1997). On the other hand, LFQPOs can also remain
relatively stable and persistent.  For example, in the case of
GRS~1915+105, a 2.0--4.5 Hz QPO is evident in every one of the 30 \xte
observations conducted over a 6-month interval in 1996--1997 (Muno et
al.\ 2001).  This degree of stability suggests that LFQPOs are tied to
the flow of matter in the accretion disk.  However, the frequencies of
these QPOs are much lower than the Keplerian frequencies for the inner
disk. For example, a 3 Hz orbital frequency around a Schwarzschild BH
of 10~\msun~corresponds to a radius near 100 $R_{\rm g}$, while the
expected range of radii for X-ray emission in the accretion disk is
\lesssim 10~$R_{\rm g}$.

LFQPOs are seen in the SPL state, the hard:SPL intermediate state, and
in some hard states, particularly when the X-ray luminosity is high
(McClintock \& Remillard 2006; Rossi et al.\ 2004). The rms amplitude
generally peaks at photon energies \gtrsim 10 keV (e.g., Vignarca et
al.\ 2003; Rodriguez et al.\ 2002), and detections have been made at
energies above 60 keV (Tomsick \& Kaaret 2001).  This behavior clearly
ties LFQPOs to the nonthermal component of the X-ray spectrum.
However, in principle, the oscillation could still originate in the
accretion disk if the PL mechanism is inverse Compton scattering of
disk photons and the coherence of the original oscillation is not
destroyed by the scattering geometry. Alternatively, the mechanism
that creates the energetic electrons required for Comptonization could
be an oscillatory type of instability.  In this case, the disk
temperature or the thermal energy flux might control the QPO
properties.

\begin{figure}[ht]           
\centerline{\psfig{figure=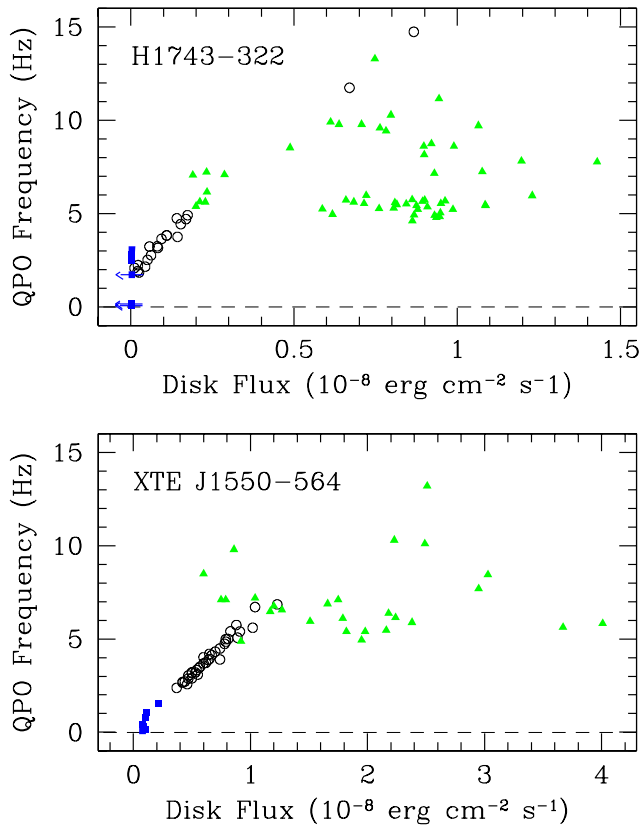,height=5in}}
\caption{Quasi-periodic oscillation (QPO) frequency versus disk flux
for XTE~J1550--564 and H~1743--322.  See Figure~\ref{fig:xo1655} for
definitions of the symbol types.  The QPO frequencies in the hard and
especially the hard:SPL intermediate state are highly correlated with
disk flux; these are C-type low-frequency QPOs (LFQPOs) based on their
phase lags (see text).  The steep power law QPOs are not correlated
with disk flux, and these are A and B type LFQPOs.}
\label{fig:freq_flux}
\end{figure}

These concepts regarding QPO origins motivate correlation studies that
compare QPO frequencies and amplitudes with various spectral
parameters.  In Figure~10 we show an example of such a correlation
between LFQPO frequency and disk flux for two sources.  The X-ray
states are symbol coded, using the conventions adopted in \S5.  The
QPO frequencies in the hard and intermediate states are highly
correlated with disk flux, but this is not true for the QPOs at higher
frequency in the SPL state.  Presently, there is no explanation for
this result.  Possibly, the QPO mechanisms could differ between the
hard and SPL states, or there could be a common mechanism that
exhibits some type of dynamical saturation as the source moves into
the SPL state.

Another avenue for QPO investigations is the study of phase lags and
coherence functions that compare two different energy bands, e.g.,
2--6 versus 13--30 keV. Such analyses have been conducted for the QPOs
in XTE~J1550-564.  Unexpectedly, the phase lag measurements showed
groups distinguished by positive, negative, and $\sim$zero lags
defining, respectively, LFQPO types A, B, and C (Wijnands et al.\
1999; Cui et al.\ 2000b; Remillard et al.\ 2002b; Casella et al.\
2004).  These details are surely complicated, but the ramifications
are very significant. The A and B types are associated with the SPL
state and the presence of high-frequency QPOs.  On the other hand,
C-type LFQPOs mostly occur in the intermediate and hard states, and
they are responsible for the frequency versus disk flux correlation
shown in Figure~\ref{fig:freq_flux}. There are also similarites
between these BH LFQPO subtypes and the three QPOs of Z-type
neutron-star systems (Casella, Belloni, \& Stella 2005).

Given the relatively high amplitudes of all LFQPOs above 6 keV, it is
clear that the C-type oscillations are well connected to both the
thermal and PL components in the X-ray spectrum.  In this sense, QPOs
can provide insights regarding the origin of the PL
spectrum. Futhermore, we now have a comprehensive archive of accurate
LFQPO measurements for a wide range of disk conditions.  These data
are available for testing any detailed models that are proposed to
explain the nonthermal states in BHBs.

\subsection{High-Frequency Quasi-Periodic Oscillations}

High-frequency QPOs (HFQPOs; 40--450 Hz) have been detected in seven
sources (5 BHBs and 2 BHCs).  These oscillations are transient and
subtle ($a \sim 0.01$), and they attract interest primarily because
their frequencies are in the expected range for matter in orbit near
the ISCO for a $\sim10$ \msun~BH.

As an aside, we briefly note that broad power peaks ($Q < 1$) have
been reported at high frequencies in a few cases (e.g., Cui et al
2000a; Homan et al. 2003; Klein-Wolt, Homan, \& van der Klis 2004). At
the present time, these broad PDS features do not impact the field
significantly, because they are relatively rare and poorly
understood. Consequently, we do not consider these broad features
further.

\begin{figure}[ht]        
\centerline{\psfig{figure=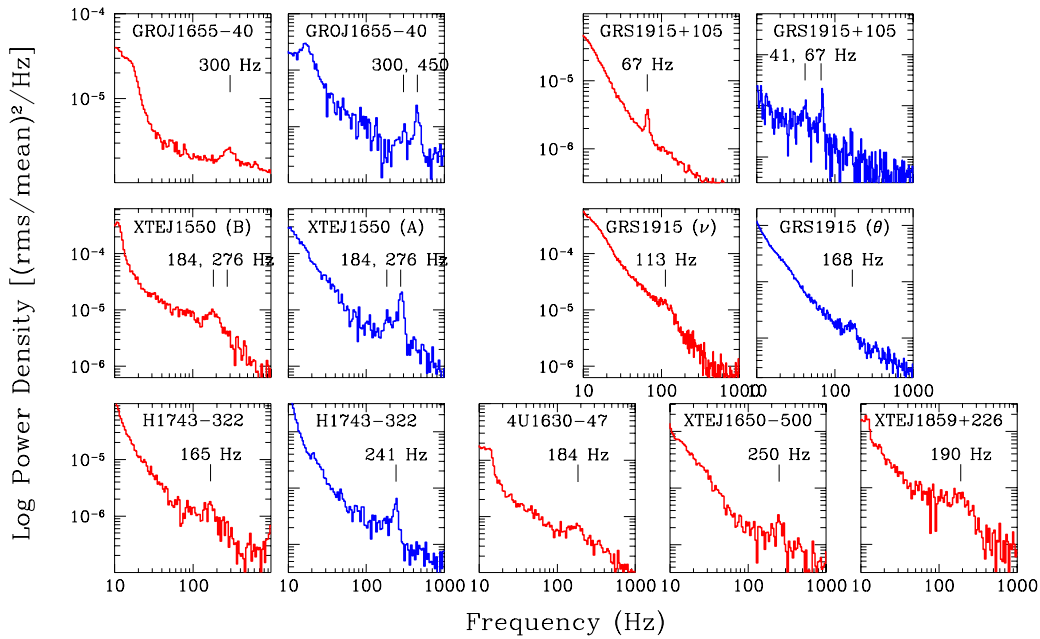,height=3.5in,angle=270.}}
\caption{High-frequency quasi-periodic oscillations (HFQPOs) observed
in black-hole binary and black-hole candidate systems.  The traces in
blue show power density spectra (PDSs) for the range 13--30 keV.  Red
traces indicate PDSs with a broader energy range, which may be either
2--30 or 6--30 keV.}
\label{fig:chfqpo}
\end{figure}

The entire sample of HFQPOs with strong detections ($ > 4 \sigma$) is
shown in Figure~\ref{fig:chfqpo}.  Three sources have exhibited single
oscillations (Cui et al.\ 2000a; Homan et al.\ 2003; Remillard 2004).
The other four sources display pairs of HFQPOs with frequencies that
scale in a 3:2 ratio. Most often, these pairs of QPOs are not detected
simultaneously. The four sources are GRO~J1655--40 (300, 450 Hz;
Remillard et al.\ 1999;  2002a; Strohmayer 2001a; Homan et al.\ 2005a),
XTE~J1550--564 (184, 276 Hz; Remillard et al.\ 2002a; Homan et al.\
2001; Miller et al.\ 2001), GRS~1915+105 (113, 168 Hz; Remillard
2004), and H~1743--322 (165, 241 Hz; Homan et al.\ 2005b; Remillard et
al.\ 2006b).  GRS~1915+105 also has a second pair of HFQPOs with
frequencies that are not in a 3:2 ratio (41, 67 Hz; Morgan et al.\
1997; Strohmayer 2001b).

HFQPOs are of further interest because they do not shift freely in
frequency in response to sizable luminosity changes (factors of 3--4;
Remillard et al.\ 2002a; 2006b).  There is evidence
of frequency shifts in the HFQPO at lower frequency (refering to the
3:2 pairing), but such variations are limited to 15\% (Remillard et
al. 2002a; Homan et al. 2005a). This is an important difference
between these BHB HFQPOs and the variable-frequency kHz QPOs seen in
accreting neutron stars, where both peaks can shift in frequency by a
factor of two (van der Klis 2005). Overall, BHB HFQPOs appear to be a
stable and identifying ``voice-print'' that may depend only on the
mass and spin of the BH (\S8.2.4).

All of the strong detections ($ > 4 \sigma$) above 100 Hz occur in the
SPL state.  In three of the sources that exhibit HFQPOs with a 3:2
frequency ratio, the $2 \nu_0$ QPO appears when the PL flux is very
strong, whereas $3 \nu_0$ appears when the PL flux is weaker (Remillard
et al.\ 2002a; 2005b).  Currently, there is no explanation for this
result.

The commensurate frequencies of HFQPOs suggests that these
oscillations are driven by some type of resonance
condition. Abramowicz and Kluzniak (2001) proposed that orbiting blobs
of accreting matter could generate the harmonic frequencies via a
resonance between a pair of the coordinate frequencies given by
GR. Earlier work had used GR coordinate frequencies and associated
beat frequencies to explain fast QPOs in both neutron-star and BH
systems (Stella et al.\ 1999), but without invoking a resonance
condition.  Current work on resonances as a means of explaining HFQPOs
includes more realistic models for fluid flow in the Kerr metric.
Resonance models are considered in more detail in \S8.2.4.

\section{Physical Models for X-ray States}

We briefly describe some physical models for the three active emission
states (\S4) and QPO phenomena (\S6).  Our focus is on basic principles
and the current interface between observations and theory.
 
\subsection{Thermal State}
 
For the thermal state there is a satisfactory paradigm, namely,
thermal emission from the inner regions of an accretion disk.
Observations and magnetohydrodynamic (MHD) simulations continue to
increase our understanding of accretion disks.  Also,
fully-relativistic models of disk spectra have recently become
publicly available, and results from this advance are discussed below.

The best-known hydrodynamic model of a radiating gas orbiting in the
gravitational potential of a compact object is the steady-state, thin
accretion disk model (Shakura \& Sunyaev 1973; Pringle 1981).  A
central problem for this model is a prescription for the viscosity
that is required to drive matter inward and heat it, while
transporting angular momentum outward.  Initially, the viscosity was
modeled using an ad hoc scaling assumption (Shakura \& Sunyaev 1973).
This model leads to a temperature profile $T(R) \propto R^{-3/4}$ and
the conclusion that the inner annulus in the disk dominates the
thermal spectrum, because $2 \pi R dR~\sigma T^4 \propto L(R) \propto
R^{-2}$.  This result has a striking observational consequence: X-ray
astronomy is the window of choice for probing strong gravity near the
horizon of an accreting stellar-mass BH.

The cardinal importance of the inner disk region highlights the need
for an accurate model for the radiation emitted near the inner disk
boundary associated with the ISCO (see \S1). The ISCO lies at $6
R_{\rm g}$ for a Schwarzschild BH ($a_* = 0$), decreasing toward $1
R_{\rm g}$ as $a_*$ approaches 1. Observationally, the thermal-state
spectra of BHBs are well fitted using the classical model for a
multitemperature accretion disk (Mitsuda et al.\ 1984; Makishima et
al.\ 1986; Kubota \& Makishima 2004; Kubota et al.\ 2005).  However,
the derived spectral parameters (i.e., the temperature and radius of
the inner disk) cannot be interpreted literally for several reasons.
The model neglects the physically-motivated torque-free boundary
condition at the ISCO (see Gierlinski et al.\ 2001; Zimmerman et al.\
2005).  Furthermore, the classical model ignores the sizable effects
due to GR and radiative transfer (e.g., see Zhang, Cui, \& Chen 1997).
Fortunately, accretion disk models for Kerr BHs have recently become
publicly available (Li et al.\ 2005; Dov\v{c}iak et al.\ 2005), and
there now exists a fully-relativistic treatment of the effects of
spectral hardening (Davis et al.\ 2005) .  Applications of these
models are discussed in \S8.

In parallel with these developments, MHD simulations have advanced our
understanding of the nature of viscosity in accretion disks.  The
magnetorotational instability (MRI) has been shown to be a source of
turbulent viscosity (Balbus \& Hawley 1991), a result that has been
confirmed by several global GR MHD simulations (e.g., DeVilliers et al.\ 2003;
McKinney \& Gammie 2004; Matsumoto et al.\ 2004).
Investigators are now considering how MRI and MHD turbulence
influences disk structure (e.g., Gammie 2004), the emerging thermal
spectrum (e.g., Merloni 2003), and Comptonization effects (e.g.,
Socrates et al.\ 2004). MHD simulations of accretion disks are
currently three-dimensional, global and based on the Kerr metric.  They will soon
include dissipative processes (i.e., radiation; see Johnson \& Gammie
2003), and it is hoped that they will then connect more directly with
observation.

\subsection{Hard State}

The association of the hard state with the presence of a steady
radio jet marked a substantial advance.  Indirect signatures of this jet
can now be recognized in the X-ray data (\S\S4--5).  However, the
relationship between the disk and jet components and the origin of the
X-ray properties of the hard state remain uncertain.

Difficulties in understanding the hard state are illustrated by
results obtained for XTE~J1118+480, a BHB with an extraordinarily
small interstellar attenuation (e.g., only 30\% at 0.3 keV) and a
display of weak outbursts confined to the hard state.  This source
provides the best direct determination of the apparent temperature and
radius of the inner disk in the hard state.  Using simultaneous HST,
EUVE, and {\it Chandra} observations (McClintock et al.\ 2001), the
disk was found to be unusually large ($\sim 100 R_{\rm g}$) and cool
($\sim 0.024$~keV).  Slightly higher temperatures
($\approx$~0.035--0.052~keV) were inferred from observations with
BeppoSAX (Frontera et al.\ 2003).  Though it seems clear that the
blackbody radiation is truncated at a large radius, the physical
condition of material within this radius remains uncertain.
Alternative scenarios include a thermal advection-dominated accretion
flow (ADAF; Esin et al.\ 2001), a radiative transition to synchrotron
emission in a relativistic flow that is entrained in a jet (Markoff et
al.\ 2001), and a radiative transition to a Compton corona (Frontera
et al.\ 2003), which must then be sufficiently optically thick to mask
the $\sim$1~keV thermal component normally seen from the disk.
Such a corona might be a hot wind leaving the disk (Blandford \&
Begelman 1999; 2004).

Recent investigations of other BHBs in the hard state suggest that
both synchrotron and Compton components contribute to the broadband
spectrum, with the Compton emission presumed to originate at the base
of the jet (Kalemci et al. 2005; Markoff, Nowak, \& Wilms 2005).  It
is also possible that the jet is supplied by hot gas from a
surrounding ADAF flow (Yuan, Cui, \& Narayan 2005).

Guidance in understanding the accretion flow in this inner region may
eventually come from other types of investigations, such as the study of
correlated optical/X-ray variability (Malzac et al.\ 2003).  Also
promising are spectral analyses that focus on features indicating
densities higher than that expected for an optically thin flow, such as
the ADAF mentioned above. One such feature is the broad Fe emission line
(e.g., Miller et al.\ 2002b; 2002c). The Fe line profile can provide
information on the Keplerian flow pattern and constrain the inner disk
radius (see \S8.2.3).  Another diagnostic spectral feature is an X-ray
reflection component (Done \& Nayakshin 2001).  In one study of the
reflection component of Cyg X--1, the hard-state disk appeared to be
truncated at a few tens of Schwarzschild radii (Done \& Zycki 1999).

\subsection{Steep Power Law State}

The physical origin of the SPL state remains one of the
outstanding problems in high-energy astrophysics.  It is crucial that
we gain an understanding of this state, which is capable of generating
HFQPOs, extremely high luminosity, and spectra that extend to
$\gtrsim$1~MeV.
 
Most models for the SPL state invoke inverse Compton scattering as the
operant radiation mechanism (see Zdziarski \& Gierli{\'n}ski
2004). The MeV photons suggest that the scattering occurs in a
nonthermal corona, which may be a simple slab operating on seed
photons from the underlying disk (e.g., Zdziarski et al.\ 2005).
Efforts to define the origin of the Comptonizing electrons have led to
models with more complicated geometry and with feedback mechanisms,
such as flare regions that erupt from magnetic instabilities in the
accretion disk (Poutanen \& Fabian 1999). There are alternative models
of the SPL state. For example, bulk motion Comptonization has been
proposed in the context of a converging sub-Keplerian flow within $50
R_{\rm g}$ of the BH (Titarchuk \& Shrader 2002; Turolla et al.\
2002).
 
An analysis of extensive \xte spectral observations of
GRO~J1655--40 and XTE~J1550--564 shows that as the PL
component becomes stronger and steeper, the disk luminosity and radius
appear to decrease while the temperature remains high.  These results
can be interpreted as an observational confirmation of strong
Comptonization of disk photons in the SPL state (Kubota \& Makishima
2004).  

\subsection{QPO Mechanisms}

In addition to spectral observations, it is also necessary to explain
timing observations of LFQPOs (0.1--30 Hz), which are commonly seen in
the SPL state. There are now a large number of proposed LFQPO
mechanisms in the literature, and we mention only a few examples here.
The models are driven by the need to account for both the observed
range of frequencies and the fact that the oscillations are strongest
at photon energies above 6 keV, i.e., where the PL component
completely dominates over the disk component.  The models include
global disk oscillations (Titarchuk \& Osherovich 2000), radial
oscillations of accretion structures such as shock fronts (Chakrabarti
\& Manickam 2000), and oscillations in a transition layer between the
disk and a hotter Comptonizing region (Nobili et al.\ 2000). Another
alternative, known as the accretion--ejection instability model,
invokes spiral waves in a magnetized disk (Tagger \& Pellat 1999) with
a transfer of energy out to the radius where material corotates with
the spiral wave. This model thereby combines magnetic instabilities
with Keplerian motion to explain the observed QPO amplitudes and
stability.

The behavior of the SPL state is complex and challenging.
Nevertheless, we have much to work with, such as the exquisite quality
of the data for this (usually) bright state, the regularities in
behavior among various sources (Figures 4--9), and the remarkable
couplings between the timing and spectral data (e.g., Figure 10).  It
appears that a successful model must allow for a highly dynamical
interplay between thermal and nonthermal processes and involve
mechanisms that can operate over a wide range of luminosity.  Finally,
we note that a physical understanding of the SPL state is required as
a foundation for building any complete model of the HFQPO
mechanisms, a topic considered in further detail in \S8.2.4.

\section{Accreting Black Holes as Probes of Strong Gravity}

The continuing development of gravitational wave astronomy is central
to the exploration of BHs.  In particular, we can reasonably
expect that LIGO and LISA will provide us with intimate knowledge
concerning the behavior of space-time under the most extreme
conditions.  Nevertheless, gravitational wave detectors are unlikely
to provide us with direct information on the formation of relativistic
jets, on strong-field relativistic MHD accretion flows, or on the
origin of high-frequency QPOs or broadened Fe lines.  Accreting BHs --
whether they be stellar-mass, supermassive or intermediate mass --
promise to provide detailed information on all of these topics and
more.  In short, accreting BHs show us uniquely how a BH interacts
with its environment.  In this section, we first sketch a scenario for
the potential impact of BHBs on physics, and we then discuss a current
frontier topic, namely, the measurement of BH spin.

\subsection{Black Holes Binaries: The Journey from Astrophysics to Physics}

Astrophysics has a long history of impacting physics: e.g., Newton's and
Einstein's theories of gravity, the ongoing research on dark matter and
dark energy, the equation of state at supranuclear densities, and the
solar neutrino puzzle.  Likewise, astrophysical BHs have the potential
to revolutionize classical BH physics; after all, the only real BHs we
know, or are likely ever to know, are astrophysical BHs.  But how can
BHBs contribute to the study of BH physics?  Very roughly, we envisage a
five-stage evolutionary program that is presently well underway.
 
{\bf Stage I}{\it ---Identify Dynamical BH Candidates}: This effort is
already quite advanced (see \S1 and Table~1) and represents an important
step because mass is the most fundamental property of a BH.
Nevertheless, the dynamical data do not probe any of the effects of strong
gravity, and therefore we curtail the discussion of Stage I.
 
{\bf Stage II}{\it ---Confirm that the Candidates are True Black
Holes}: Ideally, in order to show that a dynamical BH candidate (i.e.,
a massive compact object) is a genuine BH, one would hope to
demonstrate that the candidate has an event horizon -- the defining
characteristic of a BH.  Strong evidence has been obtained for the
reality of the event horizon from observations that compare BHBs with
very similar neutron-star binaries.  These latter systems show
signatures of the hard surface of a neutron star that are absent for
the BH systems. For example, X-ray observations in quiescence show
that the BH systems are about 100 times fainter than the nearly
identical neutron-star binaries (Narayan et al.\ 1997; Garcia et
al. 2001).  The ADAF model (\S7) provides a natural explanation for
the faintness of the BHs, namely, the low radiative efficiency of the
accretion flow allows a BH to ``hide'' most of its accretion energy
behind its event horizon (e.g, Narayan et al.\ 2002).  In quiescence,
one also observes that BHs lack a soft thermal component of emission
that is very prevalent in the spectra of neutron stars and can be
ascribed to surface emission (McClintock, Narayan, \& Rybicki 2004).  During
outburst, the presence of a surface for an accreting neutron star
likewise gives rise to distinctive phenomena that are absent in BHBs:
($a$) type I thermonuclear bursts (Narayan \& Heyl 2002; Tournear et
al. 2003; Remillard et al.\ 2006a), ($b$) high-frequency timing noise
(Sunyaev \& Revnivtsev 2000), and ($c$) a distinctive spectral component
from a boundary layer at the stellar surface (Done \& Gierli{\'n}ski
2003).

Of course, all approaches to this subject can provide only indirect
evidence for the event horizon because it is quite impossible to detect
directly any radiation from this immaterial surface of infinite
redshift.  Nevertheless, barring appeals to very exotic physics, the
body of evidence just considered makes a strong case that dynamical BH
candidates possess an event horizon.

{\bf Stage III}{\it ---Measure the Spins of Black Holes}: An
astrophysical BH is described by two parameters, its mass $M$ and its
dimensionless spin parameter $a_*$.  Because the masses of 20 BHs have
already been measured or constrained (see Stage I and \S2), the next
obvious goal is to measure spin. Indeed, several methods to measure
spin have been described in the literature, and various estimates of
$a_*$ have been published, although few results thus far can be
described as credible.  We consider this stage to be a central and 
active frontier in BH research.  Consequently we return to this
subject below, where we discuss four approaches to measuring spin and
some recent results for two BHBs.

{\bf Stage IV}{\it ---Relate Black Hole Spin to the Penrose Process
and Other Phenomena}: A number of phenomena observed in astrophysical
BHs have been argued to be associated with BH spin.  The most notable
examples are the explosive and relativistic radio jets associated with
the hard-to-soft X-ray transition that occurs near the jet line
(\S4.3).  Such ejections have been observed for at least eight BHBs
and BHCs (Mirabel \& Rodriguez 1999; Fender \& Belloni 2004;
McClintock \& Remillard 2006 and references therein). Also,
large-scale relativistic X-ray jets have been reported for
XTE~J1550-564 (Hannikainen et al. 2001; Corbel et al.\ 2002) and
H~1743--322 (Corbel et al.\ 2005).  For many years, scientists have
speculated that these jets are powered by BH spin via something like
the Penrose (1969) process, which allows energy to be milked from a
spinning BH.  Detailed models generally invoke magnetic fields
(Blandford \& Znajek 1977; Hawley \& Balbus 2002; Meier 2003; McKinney
\& Gammie 2004).  A number of beautiful ideas have been published
along these lines, but there has been no way of testing or confirming
them.  Recently, however, some progress has been made on measuring BH
spin, and it may soon be possible to attack the
jet-spin/Penrose-process connection in earnest.

{\bf Stage V}{\it ---Carry out Quantitative Tests of the Kerr Metric:}
One of the most remarkable predictions of BH physics is that the
space-time surrounding a stationary rotating BH is described by the Kerr
metric, which is completely specified by just two parameters, $M$ and
$a_*$.  Testing this prediction is the most important contribution
astrophysics can make to BH physics.  Obviously, in order to carry out
such a test, one must first measure $M$ and $a_*$ with sufficient
precision (Stage III).  Once suitable measurements of $M$ and $a_*$
have been amassed for a number of BHs, we presume that astronomer's will
be strongly motivated to devise ways of testing the metric, a topic
which is beyond the scope of this work.

\subsection{Measuring Black Hole Spin: A Current Frontier}
 
We now elaborate on Stage III by discussing four avenues for measuring
BH spin.  These include ($a$) X-ray polarimetry, which appears very
promising but thus far has not been incorporated into any contemporary
X-ray mission; ($b$) X-ray continuum fitting, which is already producing
useful results; ($c$) the Fe K line profile, which has also yielded
results, although the method is hampered by significant uncertainties;
and ($d$) high-frequency QPOs, which arguably offer the most reliable
measurement of spin once a model is established.  We now consider each
of these in turn.
 
\subsubsection{Polarimetry}
 
As pointed out by Lightman \& Shapiro (1975) and Meszaros et
al. (1988), polarimetric information (direction and degree) would
increase the parameter space used to investigate compact objects from
the current two (spectra and time variability) to four independent
parameters that models need to satisfy.  Such constraints are likely
to be crucial in our attempts to model the hard state with its radio
jet and the SPL state.  However, because of the complexities of the
accretion flows associated with these states (\S4, \S5, \& \S7) it
appears unlikely that their study will soon provide quantitative
probes of strong gravity.  We therefore focus on disk emission in the
thermal state.

The polarization features of BH disk radiation can be affected strongly
by GR effects (Connors et al.\ 1980).  The crucial requirement for a
simple interpretation is that higher energy photons come from smaller
disk radii, as they are predicted to do in conventional disk models
(\S7).  If this requirement is met, then as the photon energy increases
from 1 keV to 30 keV, the plane of linear polorization swings smoothly
through an angle of about $40^{\rm o}$ for a 9\msun~Schwarzschild BH and
$70^{\rm o}$ for an extreme Kerr BH (Connors et al.\ 1980).  The
effect is due to the strong gravitational bending of light rays.  In the
Newtonian approximation, on the other hand, the polarization angle does
not vary with energy, except for the possibility of a sudden $90^{\rm
o}$ jump (Lightman \& Shapiro 1976).  Thus, a gradual change of the
plane of polarization with energy is a purely relativistic effect, and
the magnitude of the change can give a direct measure of $a_*$.

A model is now available in XSPEC that allows one to compute the
Stokes parameters of a polarized accretion disk spectrum (Dov\v{c}iak
et al.\ 2004). While the theoretical picture is bright, and very
sensitive instruments can be built (e.g., Kaaret et al.\ 1994; Costa
et al.\ 2001), unfortunately, results to date are meager and there are
no mission opportunities on the horizon.  Important advances in this
promising area could be made by a relatively modest mission given that
BHBs in the thermal state are bright.

\subsubsection{Continuum Fitting}

Pioneering work in fitting the spectrum of the X-ray continuum to
measure spin was carried out by Zhang et al. (1997a), and the method
was advanced further by Gierli{\'n}ski et al. (2001).  Very recently,
two developments have allowed this method to be applied more widely
and with some confidence, namely: (1) models of thin accretion disks
are now publicly available in XSPEC (``KERRBB'', Li et al.\ 2005;
``KY'', Dov\v{c}iak et al. 2004) that include all relativistic effects
and allow one to fit for $a_*$; and (2) sophisticated disk atmosphere
models now exist for computing the spectral hardening factor, $f_{\rm
col} = T_{\rm col}/T_{\rm eff}$ as a function of the Eddington-scaled
luminosity of the disk (Davis et al.\ 2005).

This method of measuring $a_*$ depends on the properties of thin
accretion disks described in \S7 and is most convincing when it is
applied to BHBs in the thermal state (\S4, \S5, \& \S7).  Effectively,
in this technique, one determines the radius $R_{\rm in}$ of the inner
edge of the accretion disk and assumes that this radius corresponds to
$R_{\rm ISCO}$ (see \S1).  Because $R_{\rm ISCO}/R_{\rm g}$ is a
monotonic function of $a_*$, a measurement of $R_{\rm in}$ and $M$
directly gives $a_*$.  Provided that ($a$) $i$ and $D$ are sufficiently
well known, ($b$) the X-ray flux and spectral temperature are measured
from well calibrated X-ray data in the thermal state, and ($c$) the disk
radiates as a blackbody, it is clear that $R_{\rm in}$ can be
estimated.  A major complication is that the disk emission is not a
true blackbody but a modified blackbody with a spectral hardening
factor \fcol.  Therefore, the observations only give the quantity
$R_{\rm in}/f_{\rm col}^2$, and one needs an independent estimate
of \fcol~in order to estimate $a_*$ (Shimura \& Takahara 1995; Davis
et al.\ 2005).  A second caveat is that the orbital inclination may
differ significantly from the inclination of the BH's spin axis
(Maccarone 2002).

Using the new disk models mentioned above, Shafee et al. (2006) fitted
ASCA and \xte spectral data on the BHB GRO~J1655--40 and
found $a_* \sim 0.65-0.75$.  For 4U~1543--47, they found $a_* \sim
0.75-0.85$, although this result is based only on \xte data.
The authors consider it unlikely that either BH has a spin close to
the theoretical maximum, $a_*=1$. On the other hand, in the case
of 4U~1543--47, the estimated spin appears too high to be explained by
spin up due to accretion, which suggests that their measurements are
sensitive to the BH's natal spin.

\subsubsection{Fe K line profile}

The first broad Fe~K$\alpha$ line observed for either a BHB or an AGN
was reported in the spectrum of Cyg X--1 based on EXOSAT data
(Barr et al.\ 1985).  Since then, the line has been widely studied in
the spectra of both BHBs and AGN.  The Fe~K fluorescence line is
thought to be generated through the irradiation of the cold
(weakly-ionized) disk by a source of hard X-rays (likely an
optically-thin, Comptonizing corona).  Relativistic beaming and
gravitational redshifts in the inner disk region can serve to create an
asymmetric line profile (for a review, see Reynolds \& Nowak 2003).

The line has been modeled in the spectra of several BHBs.  In
some systems the inner disk radius deduced from the line profile is
consistent with the $6R_{\rm g}$ radius of the ISCO of a Schwarzschild
BH, suggesting that rapid spin is not required (e.g., GRS~1915+105,
Martocchia et al.\ 2002; V4641 Sgr, Miller et al.\ 2002a).  On the other
hand, fits for GX~339--4 indicate that the inner disk likely extends
inward to $(2-3)R_{\rm g}$, implying $a_*\geq0.8-0.9$ (Miller et
al.\ 2004).  XTE~J1650--500 is the most extreme case with the inner edge
located at $\approx 2R_{\rm~g}$, which suggests nearly maximal spin
(Miller et al.\ 2002c; Miniutti et al.\ 2004).  Large values of $a_*$ have
also been reported for XTE~J1655--40 and XTE~J1550--564 (Miller et
al. 2005).  Sources of uncertainty in the method include the placement
of the continuum, the model of the fluorescing source, and the
ionization state of the disk (Reynolds \& Nowak 2003).  Also, thus far
the analyses have been done using the LAOR model in XSPEC, which fixes
the spin parameter at $a_* = 0.998$ (Laor 1991).  A reanalysis of
archival data using new XSPEC models that allow one to fit for $a_*$ may
prove useful (KY, Dov\v{c}iak et al.\ 2004; KD, Beckwith \& Done 2004).

\subsubsection{High Frequency Quasi-Periodic Oscillations}

Arguably, HFQPOs (see \S6.2) are likely to offer the most reliable
measurement of spin once the correct model is known.  Typical
frequencies of these fast QPOs, e.g., 150--450 Hz, correspond
respectively to the frequency at the ISCO for Schwarzschild BHs with
masses of 15--5 \msun, which in turn closely matches the range of
observed masses (Table~1).  As noted in \S6.2, these QPO frequencies
(single or pairs) do not vary significantly despite sizable changes in
the X-ray luminosity.  This suggests that the frequencies are
primarily dependent on the mass and spin of the BH.  Those BHs
that show HFQPOs and have well-constrained masses are the best
prospects for contraining the value of the BH spin ($a_*$).

The four sources that exhibit harmonic pairs of frequencies in a 3:2
ratio (see \S6.2) suggest that HFQPOs arise from some type of
resonance mechanism (Abramowicz \& Klu{\'z}niak 2001; Remillard et
al.\ 2002a).  Resonances were first discussed in terms of specific
radii where particle orbits have oscillation frequencies in GR (see
Merloni et al. 1999) that scale with a 3:1 or a 3:2 ratio. Current
resonance concepts now consider accretion flows in a more realistic
context.  For example, the ``parametric resonance'' concept
(Abramowicz et al. 2003; Kluz\'niak et al.\ 2004; T\"or\"ok et al.\
2005) describes oscillations rooted in fluid flow where there is a
coupling between the radial and polar coordinate frequencies.  As a
second example, one recent MHD simulation provides evidence for
resonant oscillations in the inner disk (Kato 2004); in this case,
however, the coupling relation involves the azimuthal and radial
coordinate frequencies. If radiating blobs do congregate at a resonance radius
for some reason, then ray tracing calculations have shown
that GR effects can cause measurable features in the X-ray
power spectrum (Schnittman \& Bertschinger 2004).

Other models utilize variations in the geometry of the accretion flow.
For example, in one model the resonance is tied to an asymmetric
structure (e.g., a spiral wave) in the inner accretion disk (Kato 2005).
In an alternative model, state changes are invoked that thicken the
inner disk into a torus; the normal modes (with or without a resonance
condition) can yield oscillations with a 3:2 frequency ratio (Rezzolla
et al.\ 2003; Fragile 2005).  All of this research is still in a
developmental state, and these proposed explanations for HFQPOs are
basically dynamical models that lack radiation mechanisms and fail to
fully consider the spectral properties of HFQPOs described in \S6.2.

\begin{figure}[ht]          
\centerline{\psfig{figure=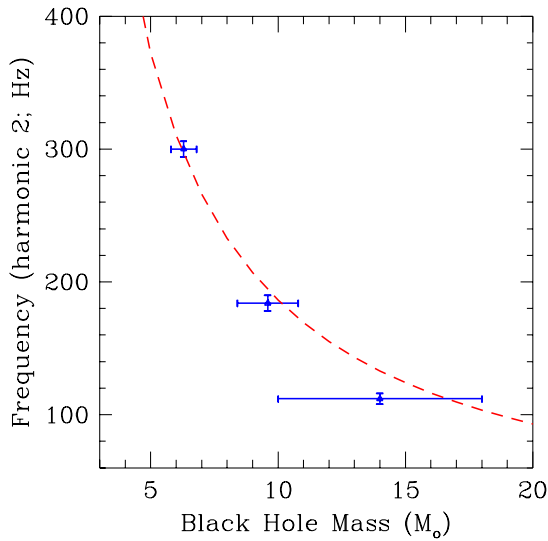,height=3.5in}}
\caption {Relationship between high-frequency quasi-periodic
oscillation (HFQPO) frequency and black hole mass for XTE~J1550--564,
GRO~J1655--40, and GRS~1915+105.  These three systems display a pair
of HFQPOs with a 3:2 frequency ratio.  The frequencies are plotted for
the stronger QPO that represents $2 \times \nu_0$. The fundamental is
generally not seen in the power spectra.  The dashed line shows a
relation, $\nu_0$ (Hz) = 931 ($M$/\msun)$^{-1}$, that fits these
data.}
\label{fig:chfqpo_m}
\end{figure}

Theoretical work aimed at explaining HFQPOs is motivated by the
following empirical result that is based on a very small sample of
three sources: XTE~J1550--564, GRO~J1655--40 and GRS~1915+105.  These
sources are presently the only ones that both exhibit harmonic (3:2)
HFQPOs and have measured BH masses.  As shown in
Figure~\ref{fig:chfqpo_m}, their frequencies appear to scale inversely
with mass (McClintock \& Remillard 2006), which is the dependence
expected for coordinate frequencies (see Merloni et al.\ 1999) or for
diskoseismic modes in the inner accretion disk (Wagoner 1999; Kato
2001).  If these HFQPOs are indeed GR oscillations, then
Figure~\ref{fig:chfqpo_m} further suggests that the three BHs have
similar values of the spin parameter $a_*$.  Obviously it is of great
importance to attempt to confirm this result by obtaining the
requisite frequency and mass measurements for additional sources.

\subsection{Critique of Methods for Measuring Spin}
 
In short, there are four avenues for measuring spin -- polarimetry,
continuum fitting, the Fe K line, and HFQPOs.  Because spin
is such a critical parameter, it is important to attempt to measure it
by as many of these methods as possible, as this will provide arguably
the best possible check on our results.  The best current method,
continuum fitting, has the drawback that its application requires
accurate estimates of BH mass ($M$), disk inclination ($i$), and
distance.  In contrast, assuming we have a well-tested model, QPO
observations require knowledge of only $M$ to provide a spin estimate.
Broadened Fe K lines and polarimetry data do not even require $M$,
although knowledge of $i$ is useful in order to avoid having to include
that parameter in the fit.  On the other hand, the Fe-line and HFQPO
methods are not well-enough developed to be applied to real systems, and
the required polarimetry data are not available, whereas the continuum
method, despite its limitations, is already delivering results.

\section{Black-Hole Binaries Yesterday, Today and Tomorrow}

Three historic milestones in stellar BH research are separated by roughly one
human generation, namely, Einstein's 1915 paper on GR, Oppenheimer and
Snyder's 1939 paper on gravitational collapse, and the 1972
identification of BH Cyg X--1 (see \S1).  Now, one more generation later, we have
measured or constrained the masses of 20 BHBs and obtained spin
estimates for two of them.  As described herein, a rich X-ray data
archive, as well as ongoing observations, are providing an intimate look
at the behavior of transient BHs as they vary in X-ray luminosity by
5--8 orders of magnitude.  We have obtained strong circumstantial
evidence for the existence of the event horizon, and observed harmonic
pairs of HFQPOs and relativistically-broadened Fe lines emanating from
near the ISCO.  However, the \xte detectors are unable to resolve the Fe
K line, and the HFQPOs are all near the limiting sensitivity of the
mission.  Further major advances will require a new timing mission with
order-of-magnitude increases in detector area, telememetry capability,
and spectral resolving power.  It is important to press ahead with a new
timing mission soon in order to complement effectively the vigorous
programs underway in gravitational wave astronomy and in observational
studies of supermassive and intermediate-mass BHs.

\section{Acknowledgements; Acronyms; Section Summaries}

\noindent {\bf Acknowledgements} \\

We are grateful to Ramesh Narayan, Jerry Orosz, Rob Fender,
and Alan Levine for contributing material for this review.  We also
thank Marek Abramowicz, Wlodek Klu{\'z}niak, Omer Blaes, Jeroen Homan,
Phil Kaaret, and Jon Miller for helpful disussions while this paper
was in preparation.  This work was funded in part by the NASA contract
to MIT for support of \xte and also by NASA Grant NNG-05GB31G to
JEM. \\ \\

\noindent {\bf Acronyms} \\

\noindent BH:   Black hole \\
BHB:  Black hole binary \\
BHC:  Black hole candidate \\
HID:  Hardness intensity diagram \\
GR:   General relativity \\
HR:   Hardness ratio \\
ISCO: Innermost stable circular orbit \\
PDS:  Power density spectrum \\
QPO:  Quasi-periodic oscillation \\
PL:   Power-law \\  \\

\noindent {\bf Summary Points} \\

\noindent 1. The topics of black-hole binaries and candidates are
introduced, along with the perspective of general relativity on black
hole mass and spin.
   
\noindent 2. The presence of a black hole is deduced from dynamical
measurements of its binary companion.  Twenty such systems are
identified.

\noindent 3. Data analysis  techniques are  summarized for  the X-ray
observations that characterize black holes.
 
\noindent 4. Accreting black holes exhibit X-ray states that are seen
as distinct and very different combinations of X-ray energy spectra
and power density spectra.

\noindent 5. The temporal evolution of X-ray states and the manner in
which states are related to primary spectral and timing properties are
illustrated for six selected sources.

\noindent 6. Quasi-periodic oscillations occur in some states. They
span a wide range in frequency, and they impose requirements on
physical models.

\noindent 7. Physical models are briefly reviewed for X-ray states and
quasi-periodic oscillations.

\noindent 8. We discuss the present and future for efforts to utilize
black-hole binaries as a test bed for applications of general
relativity.

\end{document}